\documentclass[a4paper,11pt]{article}
\pdfoutput=1 

\usepackage{jheppub} 

\usepackage{enumitem}
\usepackage{tikz}
\usetikzlibrary{decorations.pathmorphing}
\usepackage{caption}
\usepackage{tkz-tab}

\usepackage{latexsym,amssymb,amstext,amsmath}
\usepackage{hyperref}
\usepackage{graphicx} 
\usepackage{subcaption}
\usepackage{amsthm}
\usepackage{esint}
\usepackage{float}
\usepackage{color}

\def\be {\begin{equation}}
\def\ee {\end{equation}}
\def\bea {\begin{eqnarray}}
\def\eea {\end{eqnarray}}
\def\nn {\nonumber}

\def\beq{\begin{equation}}
\def\eeq{\end{equation}}
\def\beqa{\begin{eqnarray}}
\def\eeqa{\end{eqnarray}}

\newcommand{\ft}[2]{{\textstyle\frac{#1}{#2}}}

\newcommand{\sgn}{\text{sgn}}

\begin{document}
%
\begin{titlepage}

\begin{center}
 {\LARGE\bfseries 
 $R^2$ corrected AdS$_2$ holography }
 \\[10mm]

\textbf{Pedro Aniceto, Gabriel Lopes Cardoso and Suresh Nampuri }

\vskip 6mm
{\em  Center for Mathematical Analysis, Geometry and Dynamical Systems,\\
  Department of Mathematics, 
  Instituto Superior T\'ecnico,\\ Universidade de Lisboa,
  Av. Rovisco Pais, 1049-001 Lisboa, Portugal}\\
\vskip 3mm

{\tt 
pedro.aniceto@tecnico.ulisboa.pt,
gabriel.lopes.cardoso@tecnico.ulisboa.pt}, \\
{\tt  nampuri@gmail.com}
\end{center}

\vskip .2in
\begin{center} {\bf ABSTRACT } \end{center}
\begin{quotation}\noindent 
We approach the problem of constructing an explicit holographic dictionary for the AdS$_2$/CFT$_1$ correspondence
in the context of higher derivative gravitational actions in AdS$_2$ space-times. These actions are obtained by an $S^2$ reduction of four-dimensional ${\cal N}=2$ Wilsonian effective actions with Weyl squared interactions restricted to constant scalar backgrounds. BPS black hole near-horizon space-times fall into this class of backgrounds, and by identifying the boundary operators dual to the bulk fields, we explicitly show how the Wald entropy of the BPS black hole is holographically encoded in the anomalous transformation of the operator dual to a composite bulk field. Additionally, using a 2d/3d lift, we show that the CFT holographically dual to AdS$_2$ is naturally embedded in the chiral half of the CFT$_2$ dual to the AdS$_3$ space-time, and we identify the specific operator in CFT$_1$ that encodes the chiral central charge of the CFT$_2$.

\vskip 3mm
\noindent

\end{quotation}
\vfill
\today
\end{titlepage}

\tableofcontents


\section{Introduction}
\subsection{Background}
One of the most astounding insights into the nature of gravity is the holographic principle which postulates that the degrees of freedom of quantum gravity are encoded in a lower dimensional QFT. The most well fleshed out application of this principle is the celebrated AdS/CFT correspondence, which posits a description of quantum gravity in asymptotically AdS space-times in terms of a holographically dual CFT living on its 
boundary~\cite{Maldacena:1997re,Gubser:1998bc,Witten:1998qj}.
The most ubiquitous occurrence of such space-times occurs in the context of extremal black hole solutions in any gravitational theory whose near horizon geometry is, of itself, a solution to the equations of motion of the said theory and contains an AdS$_2$ factor. Hence, constructing a holographically dual description of this theory is of paramount importance to unravelling the microscopic statistical mechanical description of these black holes.
However, AdS$_2$ gravity and its holographic description offer unique challenges compared to its higher dimensional counterparts~\cite{Strominger:1998yg,Maldacena:1998uz}. These arise from the existence of two disconnected boundaries in AdS$_2$ space-times on which the putative holographically dual theory lives. This novel complication is further compounded by the peculiarity of pure AdS$_2$ gravity not supporting finite energy excitations.  Both these aspects suggest that the problem of identifying the source of the large entropy of extremal black holes is far more subtle than in the well-studied AdS$_3$/CFT$_2$ or AdS$_5$/CFT$_4$ cases. In particular, if the black hole entropy in the two-dimensional case is encoded in a dual CFT, it must arise from an exponentially large ground state degeneracy, indicating the existence of some as yet unidentified symmetry in this theory.   One way to  circumvent the problem of absence of excitations is by considering instead the nAdS$_2$/nCFT$_1$ correspondence as in \cite{Almheiri:2014cka,Maldacena:2016upp} (reviewed for example 
in~\cite{Sarosi:2017ykf}), where one obtains a dynamical model through the breaking of the conformal symmetry of the theory and perturbing away from its IR fixed point by taking a non-trivial profile for the scalar field (dilaton) coupled to the metric. Some recent examples of its application to the study of black holes are~\cite{Grumiller:2017qao,Castro2018ffi,Larsen:2018cts,Hong:2019tsx,Charles:2019tiu,Castro:2019vog}.
Nevertheless, this still leaves us with the problem of holographically determining the entropy in the constant dilaton case with unbroken conformal symmetry, corresponding to extremal RN type black holes. This was attempted in~\cite{Castro:2008ms,Castro_Song_2014ima,Grumiller:2015vaa,Cvetic:2016eiv}. In~\cite{Cvetic:2016eiv}, it was shown that the mode dual to the dilaton is an irrelevant operator, which is a non-trivial observable of the theory and transforms anomalously under asymptotic symmetry transformations. 
The associated anomalous term
was conjectured to be associated to the degrees of freedom of the theory. This still left open the question of how black hole entropy was encoded holographically, which we tackle in the current note.

\subsection{Methodology and results}
Herein, we take preliminary steps towards 
the holographic one-dimensional CFT or Conformal Quantum Mechanics (CQM) dual to quantum gravity in AdS$_2$ space-times. We focus on 
four-dimensional
4-derivative $\mathcal{N}=2$ Wilsonian  effective actions with Weyl squared ($W^2$) interactions. Such effective
actions arise in compactifications of type II superstring theory on Calabi-Yau threefolds.
These theories 
govern the dynamics of gravity coupled to Abelian gauge fields and scalars. Extremal near horizon backgrounds have constant scalar fields and curvature, determined purely in terms of the charges carried by the black hole, and hence we restrict ourselves to backgrounds with these attributes. These backgrounds factorize into $AdS_2 \times S^2$.
Evaluating the four-dimensional  action on the generic form of these backgrounds and reducing on the $S^2$ generates a 2d effective action. Restricting ourselves further to BPS configurations, we first set up an Euler-Lagrange variational principle for this action by adding suitable counterterms 
that include the Gibbons-Hawking counterterm as well as those corresponding to the gauge fields in the theory. The on-shell metric as well as the gauge fields have a radial dependence that diverges at the spatial asymptote of the AdS$_2$ space-time. Hence, the counterterms corresponding to the bulk fields will be such as to define regulated versions of these fields, which we employ as our dynamical degrees of freedom. We use the standard holographic prescription \cite{Skenderis:2002wp}  to determine the one-point function of the boundary stress tensor to be vanishing. Hence all the 2d on-shell backgrounds arising as BPS solutions of this theory correspond to ground states in the dual theory. A necessary condition that the holographically dual CFT must satisfy is that it must encode physically relevant data in 2d gravity, such as the entropy of the extremal black hole, whose near-horizon geometry gives rise to an AdS$_2$ solution of this theory. This implies the existence of an operator in the dual theory that captures the corresponding large ground state degeneracy of its vacuum. We identify the corresponding operator in the dual CFT$_1$ which encodes black hole entropy, by performing a variation of a certain composite scalar field in the bulk, ln$[ (Y^0)^2\Upsilon]$.\footnote{In the context of compactifications of type II string theory on Calabi-Yau threefolds, 
$\frac{1}{(Y^0)^2}=g^2_{top}$ is the coupling constant of the topological string theory that encodes the BPS spectrum of the type II theory, 
and $\Upsilon$ is the value of the lowest component of the  $W^2$ supermultiplet, which assumes a constant value in the near-horizon BPS backgrounds.  }
The dual operator has a vanishing one-point function in the ground state.  We then perform an asymptotic symmetry analysis.
We show that this operator transforms anomalously, and the corresponding anomaly encodes the $W^2$ corrected Wald entropy of the BPS black hole. Further, in certain cases, the 4d black hole can be lifted to a solution of 5d gravity~\cite{Gaiotto:2005gf,Castro:2007sd,Banerjee:2011ts},  in which case its near horizon geometry contains a BTZ black hole in AdS$_3$ space-time. The entropy of the 5d black hole expressed in 3d Newton's constant is precisely the BTZ entropy and it can be described as a chiral ensemble of states in the holographically dual CFT$_2$~\cite{Strominger:1997eq,Strominger:1998yg,Balasubramanian:2009bg}.
We show, following~\cite{Cvetic:2016eiv},  that the specific bulk field in 2d gravity, whose dual CFT operator encodes this BTZ 2d central charge data, is the dilaton field.  
The chiral central charge of the sector which encodes the BTZ entropy
appears in the anomaly related to asymptotic symmetry transformations of this operator. We note that the variation of the composite bulk scalar field that yields entropy data in the CQM corresponds to varying the 4d Newton's constant, while in the case of the 2d central charge the same is kept fixed. 
Based on these results we postulate that in the event of a 4d/5d lift, the CFT$_1$ holographically dual to AdS$_2$ is identified with a chiral half of the higher dimensional CFT$_2$, with the chiral 2d stress tensor expressed as a linear combination of the CFT$_1$ stress tensor as well as the dilaton operator.  
Thus, given both 2d and 1d CFT descriptions of a BPS black hole space-time, the former elegantly encodes the black hole ensemble in terms of an excitation above  the ground state of the chiral half of the CFT, while in the latter we find that the Wald entropy of the black hole is captured by the anomalous transformation of a local operator in the boundary theory.

\subsection{Organization of the paper}
This paper is organized as follows. In Section~\ref{sec_dimensionally_reduced_theory} we describe the dimensional reduction of the theory we intend to study. We complement this section with Appendix~\ref{appendix_Special_geometry}, where we provide various special geometry relations which we make use of in our dimensional reduction. In Appendix~\ref{sec:ads2_sol} we review  various parametrizations of AdS$_2$ space-times obtained by dimensional reduction.
The boundary counterterms required for the action to have a well-defined variational principle and which preserve the symplectic form on phase space are presented in Section~\ref{sec:Variational_principle}. Using those counterterms we discuss the holographic dictionary of this theory and obtain  renormalized one-point functions dual to the dynamical bulk fields, in the presence of sources. In Section~\ref{sec_PBH_analysis} we present the asymptotic symmetries of the theory and how they act on the one-point functions. 
In Section~\ref{sec:AdS2Wald} we show that for BPS black holes the Weyl squared corrected Wald entropy is encoded 
 in the anomalous transformation of the operator dual to the composite field ln $[(Y^0)^2 \Upsilon]$. 
We review three ways of computing the Wald entropy of BPS black holes in Appendix~\ref{sec:Wald_entropy}.
In Section~\ref{sec:4d5d} we review the 4d/5d lift and demonstrate that the central charge of the chiral half of the CFT$_2$, which feeds into the black hole entropy,  is captured by the anomalous transformation of the CFT$_1$ operator dual to the dilaton in the 2d gravity theory. 
We end with a summary of our main results in Section~\ref{sec_Discussion}.  
We complement this holographic approach by reviewing previous attempts~\cite{Hartman2008dq,Castro_Song_2014ima}
to recast two-dimensional gravity as a CFT  in 
Appendix~\ref{sec:HS}, wherein we derive the Dirac brackets of the theory using the Faddeev-Jackiw symplectic formalism~\cite{Faddeev:1988qp,Jackiw:1993in}.

\section{Dimensionally reduced two-dimensional bulk Lagrangian with $R^2$ interactions } \label{sec_dimensionally_reduced_theory}

There has been considerable amount of work on AdS$_2$ gravity, in conjunction with viewing the two-dimensional
theory as arising from a circle reduction of the three-dimensional Einstein-Hilbert action in the presence of a negative cosmological 
constant~\cite{Castro:2008ms,Castro_Song_2014ima,Cvetic:2016eiv,Grumiller:2017qao}. Here, we will consider a different starting point for the dimensional reduction, namely four-dimensional $\mathcal{N}=2$ supergravity theories with Weyl squared ($W^2$)
interactions. We then dimensionally reduce these theories on a two-sphere $S^2$, to obtain gravitational theories in two dimensions with AdS$_2$ vacua.
The resulting two-dimensional theories not only encode $R^2$ terms, but also exhibit electric-magnetic duality features 
inherited from the four-dimensional supergravity theories. As a consequence, the resulting two-dimensional theories are
formulated in terms of symplectic functions, and this provides a systematic way for constructing the boundary Lagrangian that needs to be added to
the two-dimensional bulk Lagrangian to ensure the existence of a well-defined variational principle.

The Wilsonian Lagrangian of the four-dimensional $\mathcal{N}=2$ supergravity theories in question describes the coupling 
of Abelian vector multiplets to $\mathcal{N}=2$ supergravity in the presence of $W^2$ interactions.\footnote{The construction
of these $\mathcal{N}=2$ supergravity theories in the framework of the superconformal multiplet calculus requires the coupling
of two so-called compensating multiplets, whose role is to ensure that the theory is gauge equivalent to a Poincar\'e supergravity
theory. One of them can be taken to be an Abelian vector multiplet, while the other can be taken to be a hyper multiplet. As shown in \cite{LopesCardoso:2000qm},
hyper multiplets play only an indirect role for the BPS backgrounds that we will consider in this paper. We have incorporated the effect of the compensating
hyper multiplet directly into the action displayed below.}
As is well-known \cite{Bergshoeff:1980is,deWit:1996gjy},
this Wilsonian Lagrangian is  encoded in a holomorphic function $F(X, \hat A)$. Here, the $X^ I$ ($I = 0, \dots, n$) denote
complex scalar fields that reside in the Abelian vector multiplets (which include one compensating multiplet), while the complex scalar $\hat A$ denotes the lowest
component of the $W^2$ supermultiplet, which is expressed in terms of the  four-dimensional anti-selfdual field strength tensor $T_{ab}^-$ as 
\bea
\hat A = (T_{ab}^-)^2 \;.
\eea
The bosonic part of the resulting four-dimensional Lagrangian can, for instance, be found in \cite{Cardoso:2006xz}. Here, we will follow the conventions of 
 \cite{Cardoso:2006xz}. This Lagrangian depends on various fields.  The set of fields of interest to us in this paper are the four-dimensional
space-time metric, the four-dimensional Abelian gauge fields $F_{\mu \nu}^I$, the complex scalar fields $X^I$ and the complex scalar $\hat A$.
The Wilsonian function $F(X, \hat A)$ is
homogeneous of second degree under scalings by $\lambda \in \mathbb{C} \backslash \{0\}$, 
\bea
F(\lambda X, \lambda^2 \hat A) = \lambda^2 F(X, \hat A) \;.
\label{FxaL}
\eea
This implies various homogeneity relations that are summarized in \eqref{sgrel}.
The function $F(X, \hat A)$ is usually given in terms of a series expansion in powers of $\hat A$,
\bea
F(X, \hat A) = F^{(0)} (X) + \sum_{g=1}^{\infty} {\hat A}^g \, F^{(g)} (X) \;.
\eea
When the coupling functions $F^{(g)}, \, g \geq 1,$ all vanish, the Wilsonian Lagrangian describes $\mathcal{N}=2$ supergravity theories without Weyl squared
interactions, and then the function $F(X, \hat A)$ equals the prepotential $F^{(0)} (X)$.

When performing the reduction of the four-dimensional theory on a two-sphere $S^2$, we take the complex scalar fields $X^I$ and $\hat A$ to be
constant, and we take the four-dimensional metric to be given by the product metric
\bea
ds_4^2 = ds_2^2 + v_2 \, d \Omega_2^2 \;,
\eea
where $ d \Omega_2^2$ denotes the line element of the two-sphere $S^2$, and 
where $v_2 >0$ denotes a constant scale factor of length dimension $2$.
In this paper we will work with two dimensionless quantities extracted from $v_2$.  The first dimensionless quantity is simply $v_2$ measured in units of the four-dimensional Newton's constant $G_4$. In the Poincar\'e gauge, $G_4$
is expressed in terms of the constant complex scalar fields as
\bea
  i \left( {\bar X}^I F_I (X, \hat{A}) - X^I {\bar F}_I ({\bar X}, \bar{\hat{A}} )\right) = G_4^{-1} \;,
\label{G4}
\eea
where $F_I (X, \hat A) = \partial F(X, \hat A)/\partial X^I$. The second dimensionless quantity is obtained by setting
\bea
v_2 = e^{-\psi} \, B^2 \;,
\eea
where $e^{-\psi}$ is a dimensionless constant field, which we refer to as the 2d dilaton,
and $B^2$ has length dimension $2$.
We then define 
the two-dimensional Newton's constant $\kappa_2^2$ by 
\bea
\frac{1}{\kappa_2^2} = \frac{B^2}{G_4} \;.
\label{G2G4}
\eea
The first dimensionless quantity is of relevance for the discussion of the entropy function \cite{Sen:2005wa,Cardoso:2006xz}.
The second dimensionless quantity is of relevance for AdS$_2$ holography \cite{Cvetic:2016eiv}.
In this section we will work with the first quantity, while in Section \ref{sec:4d5d} we will work with the 
second quantity.

We denote the local coordinates on $S^2$ by $(\theta, \varphi)$, and local
coordinates in two space-time dimensions by $(r,t)$. As mentioned above, we take the scalar fields
$X^I$ and $\hat A$ to be constant. Following \cite{Cardoso:2006xz}, 
we take $\hat A = - 4 w^2$ (with $w \in \mathbb{C}$), by setting
\begin{equation}
  \label{eq:T-w}
  T_{\underline{r}\underline{t}}{}^{-}  =- i
  T_{\underline{\theta}\underline{\varphi}}{}^{-} =
  w = constant \;,
 \end{equation}
where ${\underline{r}\underline{t}}, {\underline{\theta}\underline{\varphi}}$ denote Lorentz indices.

The four-dimensional backgrounds we consider are not only supported by constant scalar fields $X^I$ and $\hat A$,
but also by electric fields $e^I$ and magnetic charges $p^I$,
\begin{eqnarray}
ds_4^2 = ds_2^2 + v_2 \, \left( d \theta^2 + \sin^2 \theta \, d \varphi^2 \right) \;, \nonumber\\
F_{rt}^I = e^I \;\;\;,\;\;\; F_{\theta \varphi}^I = p^I\, \sin \theta \;, \nonumber\\
{\hat A} = - 4 w^2 \;\;\;,\;\;\; X^I = constant \;,
\label{backgr}
\end{eqnarray}
where the electric fields $e^I = e^I (r,t)$ are functions of $(r,t)$.
The four-dimensional Ricci scalar splits into
\begin{equation}
R_4 = R_2 - \frac{2}{v_2} \;.
\label{Ric4}
\end{equation}
Note that in the conventions used here \cite{Cardoso:2006xz}, the Ricci scalar of $S^2$ is negative.

We will focus on solutions to the four-dimensional equations of motion which are such that the two-dimensional space-time metric has constant curvature scalar,\footnote{This is motivated by the fact that we will be interested in four-dimensional backgrounds that are $AdS_2 \times S^2$.}
\bea
R_2 = \frac{2}{v_1} \;\;\;,\;\;\; v_1 > 0 \;,
\label{constcurv}
\eea
where $v_1$ denotes a constant scale factor of length dimension $2$, measured in units of $G_4$.

We denote the metric  in two space-time dimensions by
\bea
ds_2^2 =  h_{ij} dx^i dx^j \;\;\;,\;\;\; x^i = (r,t) \;.
\label{hmet}
\eea
Locally, it can be brought into Fefferman-Graham form by a bulk diffeomorphism, 
\bea
ds^2_2 &=& dr^2 + {h}_{tt} (r,t) \, dt^2  \;.
\label{FGg}
\eea
In two dimensions, the Riemann tensor only contains one independent component, namely $R_2$, and hence
\bea
R_{ij} = \frac12 R_2  \, h_{ij} \;.
\label{RRh}
\eea
In Fefferman-Graham gauge \eqref{FGg}, 
any two-dimensional metric $h_{ij}$ with constant $R_2 = 2/v_1$ is locally of the form \eqref{FGg} with \cite{Cvetic:2016eiv}
\bea
 - {h}_{tt} &=&\left(  {\alpha} (t) \, e^{r/\sqrt{v_1}} + {\beta} (t) \, e^{-r/\sqrt{v_1}} \right)^2 \;,
 \nonumber\\
 \sqrt{- h_2}  &=& {\alpha} (t) \, e^{r/\sqrt{v_1}} + {\beta} (t) \, e^{-r/\sqrt{v_1}}  \;.
\label{metfeff}
\eea
Various choices of $\alpha$ and $\beta$ yield known forms of two-dimensional metrics: the choice $\alpha =1$ and $\beta =0$
describes AdS$_2$ space-time in a Poincar\' e patch and is 
associated with the near-horizon geometry of an extremal black hole in four dimensions; 
the choice  $\alpha =1$ and $\beta =-1$ describes the near-horizon geometry
of a near-extremal black hole \cite{Sen:2011cn};
 the choice $\alpha = 1, \beta = 1$ corresponds to a smooth horizonless geometry with two AdS$_2$ boundaries  \cite{Cvetic:2016eiv}.
 We refer to Appendix \ref{sec:ads2_sol} for a brief review of some of these solutions.

Let us now discuss the evaluation of the four-dimensional Wilsonian Lagrangian  
in a background of the form \eqref{backgr} and its subsequent reduction on a two-sphere $S^2$. 
The resulting action, which we display below in \eqref{bulkr2lag2}, describes a consistent truncation of the
four-dimensional theory for all backgrounds that are supported by constant scalar fields and that are of the form \eqref{backgr}.
This reduction was done in 
\cite{Cardoso:2006xz} for the case of a two-dimensional metric of the form \eqref{metfeff} with $\alpha =1$ and $\beta =0$.
Here, we apply the results of \cite{Cardoso:2006xz} to the case of product metrics \eqref{backgr}.  The results in 
\cite{Cardoso:2006xz}
are given in terms of the rescaled fields
\begin{eqnarray}
   Y^I = \ft14 v_2 \, {\bar w} \, X^I \,,\quad
  \Upsilon = \ft{1}{16} v_2^2 \, {\bar w}^2 \, {\hat A} = - \ft14
  v_2^2 \, \vert w \vert^4  \;.
  \label{rescalXA}
  \end{eqnarray}
 Note that $Y^I$ and $\Upsilon$ are constant scalar fields, since $v_2, X^I, {\hat A}$ are taken to be constant.
 Also note that  
 $\Upsilon$ is real and negative. In the rescaled variables, the relation \eqref{FxaL} becomes
 \bea
F(\lambda Y, \lambda^2 \Upsilon) = \lambda^2 F(Y, \Upsilon) \;.
\label{FYUps}
\eea
 Then, using \eqref{G4} and \eqref{rescalXA}, we infer the relation
 \bea
 i \left( {\bar Y}^I F_I (Y, \Upsilon) - Y^I {\bar F}_I ({\bar Y}, {\bar \Upsilon} ) \right) = \frac{v_2 \,  \sqrt{-\Upsilon}}{8 G_4} \;.
  \label{relYg2g4}
 \eea

Upon reduction on $S^2$, the resulting two-dimensional bulk Lagrangian can be expressed in terms of the set of constant
scalar fields $(v_2, X^I, {\hat A})$, or equivalently, in terms of the set of constant
scalar fields $(v_2, Y^I, {\Upsilon})$.  We follow \cite{Cardoso:2006xz} and give the reduced bulk Lagrangian in terms of the latter set
of constant scalar fields. We write the reduced bulk Lagrangian as 
${\cal F} = {\cal F}_1 + {\cal F}_2$ (we refer to Appendix 
\ref{appendix_Special_geometry} for details on the reduction), where ${\cal F}_1$ contains the electric fields $e^I$ and magnetic charges $p^I$,
while ${\cal F}_2$ contains gravitational terms proportional to powers of $R_2$,
 \begin{eqnarray}
  \label{eq:F1-F2}
\ft12   \mathcal{F}_1 &=&{}  \tfrac18 N_{IJ} \Big[  ( \sqrt{-h_2}/v_2)^{-1}
  e^Ie^J - \frac{\sqrt{-h_2}}{v_2} \, {p^I} {p^J} \Big] -
  \tfrac14(F_{IJ}+ \bar F_{IJ}) {e^Ip^J}  \nonumber \\
  &&
  + \ft12 i e^I\Big[ F_I + F_{IJ} \bar Y^J   - \mathrm{h.c.}\Big] 
  - \ft12 \frac{\sqrt{-h_2}}{v_2} \,  p^I\Big[ F_I - F_{IJ} \bar Y^J  +  \mathrm{h.c.}\Big]\;,
  \nonumber\\[2mm] 
\ft12   \mathcal{F}_2 &=&  
  \frac{4 i}{\sqrt{-\Upsilon}} (\bar Y^IF_I -Y^I\bar F_I) \, \frac{\sqrt{-h_2}}{v_2} \, 
  (1 - \ft12 v_2 \, R_2 )  \nonumber\\
  &&
  + i \frac{\sqrt{-h_2}}{v_2} \,  \Big[ F-Y^IF_I - 2\Upsilon F_\Upsilon+  \tfrac12 \bar
  F_{IJ}Y^IY^J - \mathrm{h.c.} \Big] \nonumber\\
  &&
  + i (F_\Upsilon-\bar F_\Upsilon) \, \frac{\sqrt{-h_2}}{v_2} \, 
  \Big [ 8\, v_2^2 \, R_2^2 - 32 v_2 \, R_2 + 32  - 8 ( \ft12 v_2 \, R_2 +1)  \sqrt{-\Upsilon} \Big]
 \;,
\end{eqnarray}
where $F_I = \partial F(Y, \Upsilon) / \partial Y^I, \, F_{\Upsilon} = \partial  F(Y, \Upsilon) / \partial \Upsilon, \, F_{IJ} = 
 \partial^2 F(Y, \Upsilon) / \partial Y^I \partial Y^J$ and 
\begin{equation}
N_{IJ} = -i \left( F_{IJ} - \bar{F}_{IJ} \right) \;.
\end{equation}

 The two-dimensional bulk Lagrangian ${\cal F} = {\cal F}_1 + {\cal F}_2$ depends on a set of dynamical fields, namely 
 the electric fields $e^I$ and the two-dimensional 
 space-time metric $h_{ij}$, and on the constant fields $Y^I, \Upsilon, v_2$ as well as on the magnetic charges $p^I$.
 To make the underlying electric-magnetic duality manifest, we now perform a Legendre transformation  of $\cal F$ with respect to the $p^I$,
thereby replacing $p^I$ by the conjugate quantity $f_I = - \partial {\cal F} / \partial p^I $ \cite{Cardoso:2006xz}. 
Note that $f_I$ is a field strength in two dimensions, i.e.  $f_I = G_{I \, rt}$ with $G_I$ a two-form. 
The resulting Lagrangian $H(e^I, f_J) $,\footnote{We omit the dependence on the constant scalars $v_2, Y^I, {\Upsilon}$, for notational simplicity.}
\begin{equation}
H(e^I, f_J) =  {\cal F} (e^I, p^J)  + p^I f_I \;,
\end{equation}
is the two-dimensional bulk Lagrangian for which we will set up a variational principle in the next section. 
We obtain
\begin{eqnarray}
H(e^I, f_J) 
&=& \tfrac14 \, ( \sqrt{-h_2}/v_2)^{-1} \, N_{IJ} \, e^Ie^J \nonumber\\
&& + ( \sqrt{-h_2}/v_2)^{-1} \,  N^{IJ} \left( f_I - \ft12 \left(F_{IK} + {\bar F}_{IK} \right) e^K - \frac{\sqrt{-h_2}}{v_2} \,  \Big[ F_I - F_{IK} \bar Y^K  +  \mathrm{h.c.}\Big]
 \right) \nonumber\\
&& \qquad  \left( f_J - \ft12 \left(F_{JL} + {\bar F}_{JL} \right) e^L - \frac{\sqrt{-h_2}}{v_2} \,   \Big[ F_J - F_{JL} \bar Y^L  +  \mathrm{h.c.}\Big]
 \right)
  \nonumber \\
  &&
  + i e^I\Big[ F_I + F_{IJ} \bar Y^J   - \mathrm{h.c.}\Big] \nonumber\\
  &&
 +  \frac{8 i }{\sqrt{-\Upsilon}} (\bar Y^IF_I -Y^I\bar F_I) \, \frac{\sqrt{-h_2}}{v_2} \, 
  (1 - \ft12 v_2\, R_2 )  \\
  &&
  + 2  i \frac{\sqrt{-h_2}}{v_2} \Big[ F-Y^IF_I - 2\Upsilon F_\Upsilon+  \tfrac12 \bar
  F_{IJ}Y^IY^J - \mathrm{h.c.} \Big] \nonumber\\
  &&
  + 2 i (F_\Upsilon-\bar F_\Upsilon) \, \frac{\sqrt{-h_2}}{v_2} \, 
  \Big [ 8\, v_2^2 \, R_2^2 - 32 v_2 \, R_2 + 32  - 8 ( \ft12 v_2 \, R_2 +1)  \sqrt{-\Upsilon} \Big]
 \;. \nonumber
 \label{bulkr2lag}
\end{eqnarray}
Note that $H(e^I, f_J) $ appears to be singular in the limit $\Upsilon \rightarrow 0$, but this is an artefact of having chosen to give
the bulk Lagrangian in terms of the constant scalars $v_2, Y^I, {\Upsilon}$, instead of the original constant scalars $v_2, X^I, {\hat A}$.

In order to make the electric-magnetic duality of $H(e^I, f_J) $ manifest, we define
\bea
R_{IJ} = F_{IJ} + {\bar F}_{IJ} \;\;\;,\;\;\; N^{IJ} N_{JK} = \delta^I_K \;.
\eea
Then
\begin{eqnarray}
H(e^I, f_J) 
&=& \tfrac14 \, ( \sqrt{-h_2}/v_2)^{-1} \, (e^I, f_I)
\begin{bmatrix}
N_{IJ} + R_{IK} N^{KL} R_{LJ}  & \;\;\; - 2 R_{IK} N^{KJ} \\
- 2 N^{I K} R_{KJ} & 4 N^{IJ}
\end{bmatrix} \, 
\begin{pmatrix}
e^J \\ 
f_J
\end{pmatrix}
 \nonumber\\
&& +   (e^I, f_I) \left[ 2i 
\begin{pmatrix}
F_I - {\bar F}_I \\
- (Y^I - {\bar Y}^I) 
\end{pmatrix}
+ 4 \Upsilon
\begin{pmatrix}
\bar{F}_{IK} N^{KL} F_{\Upsilon L}  \\
- N^{IJ} F_{\Upsilon J} 
\end{pmatrix}
+ 4 \bar{\Upsilon} 
\begin{pmatrix}
{F}_{IK} N^{KL} \bar{F}_{\Upsilon L}  \\
- N^{IJ} \bar{F}_{\Upsilon J} 
\end{pmatrix}
\right]
\nonumber\\
&& - \sqrt{-h_2} \, P(R_2) \nonumber\\
&& + \frac{\sqrt{-h_2}}{v_2} \,  \Big\{
\frac{8\mathrm{i}}{\sqrt{-\Upsilon}} (\bar Y^IF_I -Y^I\bar F_I) \,
- 2i (\bar Y^I F_I -Y^I\bar F_I) \nonumber\\
&& \left. 
\qquad \qquad \qquad - 2i ( \Upsilon \, F_{\Upsilon} - \bar{\Upsilon} \bar{F}_{\Upsilon} ) 
+ 8 \Upsilon \bar{\Upsilon} \, \bar{F}_{\Upsilon I} N^{IJ} F_{\Upsilon J} 
\right. \nonumber\\
&&  \qquad \qquad \qquad + 2 \Upsilon F_{\Upsilon I} N^{IJ} (F_J - \bar{F}_{JL} Y^L) + 
 2 \bar{\Upsilon} \bar{F}_{\Upsilon I} N^{IJ} (\bar{F}_J - F_{JL} \bar{Y}^L) 
\nonumber\\
  &&  \qquad \qquad \qquad 
  + 2 \mathrm{i}(F_\Upsilon-\bar F_\Upsilon) \,
  \Big (  32  - 8  \sqrt{-\Upsilon} \Big)
  \Big\}
 \;,
 \label{bulkr2lag2}
\end{eqnarray}
where $P(R_2)$ denotes the following polynomial in $R_2$,
\bea
P(R_2) &=&  4 \frac{i }{\sqrt{-\Upsilon}}  (\bar Y^IF_I -Y^I\bar F_I) \,
 \, R_2
  - \frac{2 i (F_\Upsilon-\bar F_\Upsilon) }{v_2} \,
  \Big ( 8\, v_2^2  \, R_2^2 - 32 v_2 \, R_2  - 4 v_2 \, R_2 \,  \sqrt{-\Upsilon} \Big) \;, \nonumber\\
  P'(R_2) &=&   \frac{d P}{dR_2 }=  4 \frac{i }{\sqrt{-\Upsilon}}  (\bar Y^IF_I -Y^I\bar F_I) 
  - 2 i (F_\Upsilon-\bar F_\Upsilon) \,
  \Big ( 16\, v_2  \, R_2 - 32   - 4  \,  \sqrt{-\Upsilon} \Big) \;.
  \label{FFp}
\eea
Using \eqref{relYg2g4}, we exhibit the dependence on $G_4$ in these expressions,
\bea
P(R_2) &=&  
\tfrac12 \frac{v_2}{G_4}
 \, R_2 
  - \frac{2 i (F_\Upsilon-\bar F_\Upsilon) }{v_2} \,
  \Big ( 8\, v_2^2  \, R_2^2 - 32 v_2 \, R_2  - 4 v_2 \, R_2 \,  \sqrt{-\Upsilon} \Big) \;, \nonumber\\
  P'(R_2) &=&    \tfrac12 \frac{v_2}{G_4} 
 - 2 i (F_\Upsilon-\bar F_\Upsilon) \,
  \Big ( 16\, v_2  \, R_2 - 32   - 4  \,  \sqrt{-\Upsilon} \Big) \;.
\eea
Observe that in the absence of Weyl squared interactions in four dimensions, $F_{\Upsilon} = 0$, and $P(R_2)$ reduces
to the Einstein-Hilbert term $R_2$.

The first two lines of \eqref{bulkr2lag2} contain the dependence on the field strengths $e^I, f_J$. The third line
contains the terms that depend on powers of the curvature scalar $R_2$. The remaining lines contain terms that 
depend on the scalar fields $Y^ I$ and $\Upsilon$.
Each of the lines of \eqref{bulkr2lag2} is a symplectic scalar, i.e. each line transforms as a scalar under electric-magnetic
duality transformations \cite{deWit:1996gjy,LopesCardoso:2000qm,Mohaupt:2011aa,LopesCardoso:2019mlj}.\footnote{Even though $\Upsilon$ is real, we have chosen
to use both $\Upsilon$ and $\bar \Upsilon$ in these expressions and in the ones given below to make the symplectic structure manifest.}

We recall that 
another symplectic scalar function is given by
\bea
H(e^I, f_J) + q_I e^I - p^I f_I = {\cal F} (e^I,p^J) + q_I e^I \;,
\label{entrofunc}
\eea
which, when evaluated in an AdS$_2$ background \eqref{metfeff} with $\alpha = 1, \beta =0$,
becomes the entropy function for extremal black holes \cite{Cardoso:2006xz}.

In the next section, we will set up the variational principle for the $R_2^2$ corrected bulk Lagrangian \eqref{bulkr2lag2}.
In this Lagrangian, the dynamical fields are the field strengths $e^I, f_J$ and the two-dimensional 
 space-time metric $h_{ij}$.  For later use, we note the following relations,
 \bea
\frac{ \partial  H}{\partial e^I} &=& \tfrac12 \, ( \sqrt{-h_2}/v_2 )^{-1} \Big[ N_{IJ} e^J  - 2 (R N^{-1})_I{}^J 
(f_J - \tfrac12 R_{JK}  e^K)
\Big] \nonumber\\
&& + 2i (F_I - \bar{F}_I) + 4 \Upsilon \bar{F}_{IK} N^{KL} F_{\Upsilon L} + 4 \bar{\Upsilon} 
{F}_{IK} N^{KL} \bar{F}_{\Upsilon L}  \;, \nonumber\\
\frac{ \partial  H}{\partial f_I} &=& 2 ( \sqrt{-h_2}/v_2 )^{-1} \, N^{IJ} ( f_J - \tfrac12 R_{JK}  e^K) \nonumber\\ 
&& - 2 i (Y^I - \bar{Y}^I)   - 4 \Upsilon 
N^{IJ} F_{\Upsilon J}  - 4 \bar{\Upsilon}  N^{IJ} \bar{F}_{\Upsilon J} \;.
\label{derHef}
\eea
We also note
the following identity, which can be verified in a straightforward manner,
\bea
&& e^I \, \delta \left( \frac{ \partial  H}{\partial e^I} \right) + f_I \,  \delta \left( \frac{ \partial  H}{\partial f_I} \right) = 
 \tfrac14 \, ( \sqrt{-h_2}/v_2)^{-1} \, \nonumber\\
 && \qquad  \delta \left[  (e^I, f_I)
\begin{bmatrix}
N_{IJ} + R_{IK} N^{KL} R_{LJ}  &\; \;\; - 2 R_{IK} N^{KJ} \\
- 2 N^{I K} R_{KJ} & \;\;  4 N^{IJ}
\end{bmatrix} \, \begin{pmatrix}
e^J \\ 
f_J
\end{pmatrix}
\right] \;,
\label{varHef}
\eea
where the variation $\delta$ is with respect to $e^I$ and to $f_J$.

\section{Variational principle and holographic renormalization  \label{sec:Variational_principle} }

\subsection{Variational principle in the presence of $R_2^2$ interactions \label{sec:var-prin-r2}}

The bulk Lagrangian \eqref{bulkr2lag2} depends on the dynamical fields $h_{ij}, e^I , f_J$ as well as on the constant scalar fields
$v_2, Y^I , \Upsilon$. In the following, we set up a consistent variational principle for the dynamical fields,
by viewing space-time as a two-dimensional manifold $M$ with boundaries, and imposing Dirichlet boundary conditions on the dynamical fields
at these boundaries, following  the procedure of \cite{Cvetic:2016eiv}. 

For concreteness, we will assume $M$ to possess just one boundary $\partial M$, which we take to be time-like. The discussion given below can be extended
to include a second boundary associated with the  presence of a black hole horizon, say, see Appendix~\ref{sec:Wald_entropy} for more details.

Thus, we 
foliate the two-dimensional bulk space-time $M$ by a sequence of one-dimensional timelike lines homeomorphic to
the boundary $\partial M$. We take $t$ to be the coordinate along a given timelike line, and we take $r$ to be the spacelike coordinate, as in \eqref{metfeff}. 
We will refer to the timelike line at fixed $r$ as the boundary $\partial M_r$
of the interior region $M_r$. 
$\partial M$ denotes
the timelike line at $r = \infty$. 

In this paper, we will be interested in studying 
four-dimensional BPS black holes in the presence of Weyl squared interactions
 from a two-dimensional point of view. We will therefore demand that the variational principle allows for such solutions. This, in particular, 
 implies that we will focus on two-dimensional solutions based on metrics \eqref{metfeff}  with constant curvature $R_2$ satisfying the BPS condition 
 $v_1 = v_2$.

We express the field strengths $e^I = F_{rt}^I $ and $f_I = G_{I \, rt} $ in terms of gauge potential one-forms, 
\bea
e^I = F_{rt}^I = \partial_r A_t^I - \partial_t A_r^I \;\;\;,\;\;\; f_I = G_{I \, rt} = \partial_r \tilde{A}_{I \, t} - \partial_t \tilde{A}_{I \, r} \;.
\eea
First, we consider the variation of the bulk Lagrangian $H(e^I,f_J)$ with respect to the gauge potentials.
For concreteness, we focus on the variation of $H$ with respect to $A^I = A^I_i dx^i$, omiting the index $I$,
\bea
\delta H &=& \frac{\partial H}{\partial e} \, \delta e = \frac{\partial H}{\partial e}  \left( \partial_r \delta A_t - \partial_t \delta A_r \right) \nonumber\\
&=& \partial_r \left( \frac{\partial H}{\partial e}   \,  \delta A_t  \right) - \partial_t  \left( \frac{\partial H}{\partial e}   \,  \delta A_r  \right) \nonumber\\
&& - \left(  \partial_r  \frac{\partial H}{\partial e} \right) \delta A_t  + \left(  \partial_t \frac{\partial H}{\partial e} \right) \delta A_r \;.
\eea
The last line gives the equations of motion for the electric field $e$. 
The second line gives rise to two boundary terms.  To deal with the second boundary term, 
we take $t$ to lie in the interval $[ t_i, t_f ]$, and we demand that at the initial and final
points $t_i$ and $t_f$,
\bea
\delta A_r = 0 \;,
\label{delAr}
\eea
and similarly, $\delta \tilde{A}_{r} =0$.
We will enforce these conditions by working in the gauge $A^I_r =  \tilde{A}_{I \, r}  = 0 $.

To deal with the first boundary term, and with the goal of renormalizing it, 
we follow the canonical procedure detailed in \cite{Papadimitriou:2010as,Cvetic:2016eiv} and
add a boundary action to the bulk action that involves two new {\it independent} fields, namely $\pi$, which is the momentum 
conjugate to $A_t$, and $A^{ren}_t$, which will correspond to the renormalized field,
  \bea
S_{total} =  \int_M d^2x \, H(e) +  S_1   - \int_{\partial M} dt \,  \pi \,  \left( A_t  - A^{ren}_t \right) \;,
 \label{buboact}
 \eea
 where the boundary action $S_1$ depends on $\pi$, but not on $A_t$ or on $A_r$,
 \bea
 S_1 = \int_{\partial M} L_1( \pi) \;.
 \eea
 Then, varying the bulk-boundary action 
 \eqref{buboact} with respect to $A_t$,  $\pi$ and $A^{ren}_t$ results in the boundary terms
    \bea
 && \int_{\partial M} dt \, \left[  \left(  \frac{\partial H}{\partial e} - \pi \right)  \delta A_t 
  - \left( A_t - \frac{\partial L_1}{\partial \pi} - A^{ren}_t \ \right) \, \delta \pi + \pi \, \delta A^{ren}_t
 \right] \;.
 \label{bubovar}
  \eea
 The first two variations are made to vanish by demanding that at the boundary $\partial M$, $\pi$ and  $A^{ren}_t$ satisfy the on-shell
 relations
 \be
  \pi = \frac{\partial H}{\partial e} \;, \quad A^{ren}_t = A_t - \frac{\partial L_1}{\partial \pi } \;.
  \ee
  The third variation is made to vanish by imposing the Dirichlet boundary condition
  \bea
   \delta A^{ren}_t \vert_{\partial M} = 0 \;.
   \eea
 
The above discussion can be easily generalized to the case of field strengths $e^I, f_J$. The bulk-boundary action now takes the form,
\bea
S_{total} =  \int_M d^2 x \, H(e) +  S_1   - \int_{\partial M} dt \,  \pi_I \,  \left( A^I_t  - A^{ren \, I}_t \right) 
- \int_{\partial M} dt \,  {\tilde \pi}^I \,  \left( {\tilde A}_{I \, t}   - {\tilde A}^{ren}_{I \, t} \right) \;, 
\label{bbppaar2}
 \eea 
 where the boundary action $S_1$ does not depend on $A$ and $\tilde A$,
   \bea
 S_1 = \int_{\partial M} L_1 (\pi_I,  {\tilde \pi}^I )  \;.
 \eea
  We obtain a consistent variational principle by
imposing the Dirichlet boundary conditions
\bea
\delta  A^{ren \, I }_t \vert_{\partial M} = 0 \;\;\;,\;\;\;
\delta  {\tilde A}^{ren}_{I \, t}  \vert_{\partial M} = 0 \;,
\label{varAren}
\eea
and demanding that at the boundary $\partial M$, the following on-shell relations hold, 
 \be
  \pi_I = \frac{\partial H}{\partial e^I} \;, \quad {\tilde \pi}^I = \frac{\partial H}{\partial f_I} \;, \quad 
  A_t^{ ren \, I} = A_t^{ I }  - \frac{\partial L_1}{\partial \pi_I } \;, \quad  
   {\tilde A}^{ren}_{I \, t} = \tilde{A}_{ I \, t }  - \frac{\partial L_1}{\partial {\tilde \pi}^I } \;.
  \label{piArenos}
  \ee
   
Next, we consider the variation of the bulk Lagrangian $H(e^I,f_J)$ with respect to the space-time metric $h_{ij}$. We follow 
\cite{Harlow:2019yfa} and use some of their results. 
To set up the variational principle we define the unit normal $n_i$ to the timelike hypersurface such that
\be
h_{ij}= \gamma_{ij } + n_i n_j \,, 
\ee
with $\gamma_{ij }$ the induced metric on $\partial M$. Taking $k(x^i) =0$ as a function parameterizing the boundary we have
\be \label{eq_definition_normal}
n_i = \frac{\partial_i k}{\sqrt{h^{jl} \partial_j k \partial_l k  }} \,.
\ee 
With these definitions, the extrinsic curvature tensor is given by $K_{ij} = \gamma_i^{\;k} \nabla_k n_j  $. Hence, the variation of $K \equiv h^{ij}K_{ij}$ is 
\be
2 \delta K = h^{ij} n^k \nabla_k \delta h_{ij} - n^i \nabla^j \delta h_{ij} - K^{ij} \delta h_{ij} - \mathcal{D}_i \left(\gamma^{ij} n^k \delta h_{k j }  \right) \,,
\ee
with $\mathcal{D}_i $ the boundary covariant derivative defined by 
\be
\mathcal{D}_i T^{i_1 i_2 \dots}_{\qquad j_1 j_2 \dots} \equiv \gamma^{\;j}_i \gamma_{k_1}^{\; i_1} \gamma_{k_2}^{\; i_2} \dots \gamma_{j_1}^{\; l_1}  \gamma_{j_2}^{\; l_2}  \dots  \nabla_j T^{k_1 k_2\dots}_{\qquad  l_1 l_2 \dots}\,. 
\ee 
On the other hand, using\footnote{Recall that we use conventions in which the Ricci scalar of $S^2$ is negative.} $\delta R = \left(h^{ij} \Box - \nabla^i \nabla^j \right) \delta h_{ij} - R^{ij} \delta h_{ij}$ the variation of the bulk term $- \sqrt{-h_2} \, P(R_2)$ results in
\bea 
&& -\delta \int_M d^2x \sqrt{-h_2} P(R_2)  = \nonumber\\
&& \int_{M} d^2x \sqrt{-h_2} \left[ P'(R_2) {R_2}^{ij}-\frac{h^{ij}}{2} P(R_2)  + \nabla^i \nabla^j P'(R_2) - h^{ij} \Box P'(R_2)   \right] \delta h_{ij}  \nn  \\
& & + \int_{\partial M } dt \sqrt{-\gamma} \left\{  P'(R_2) \left[- 2 \delta K  - K^{ij}\delta h_{ij}  - \mathcal{D}_i  \left(\gamma^{ij} n^k \delta h_{k j }  \right)   \right] \right. \nn \\
& & \qquad \qquad \qquad  + \left[ h^{ij}   \nabla_n P'(R_2) - n^i \nabla^j P'(R_2)   \right] \delta h_{ij} \Big \} \,, 
\label{eq_variation_bulk_F_R}
\eea
where $\nabla_n = n^i \nabla_i$.
To this action we add the boundary counterterm $2 P'(R_2) K $ \cite{Dyer:2008hb}, whose variation yields
\be \label{eq_variation_GH_ct}
\delta \int_{\partial M } dt \sqrt{-\gamma} \, 2 \, P'(R_2) \, K = \int_{\partial M } dt \sqrt{-\gamma} \left[ 2P'(R_2) \left(\delta K  + \frac{K}{2} \gamma^{ij} \delta \gamma_{ij}\right)  +2 P''(R_2) K \delta R_2   \right] \,.
\ee
Using $K^{ij} n_j = 0$, it follows that $ K^{ij} \delta h_{ij} = K^{ij} \delta \gamma_{ij}$. 
Then, since in two dimensions $( K^{ij} -  K \gamma^{ij} )  \delta \gamma_{ij} = 0 $, the 
 first two boundary terms of~\eqref{eq_variation_bulk_F_R} cancel against the  first two boundary terms of~\eqref{eq_variation_GH_ct}.

Since the bulk Lagrangian $H(e^I,f_J)$ given in \eqref{bulkr2lag2} contains the second order polynomial $P(R_2)$ detailed in~\eqref{FFp}, the variation of the boundary term that we have just discussed leads to the presence of the $\delta R_2$ term in~\eqref{eq_variation_GH_ct}, which requires the addition of further boundary terms. 
To discuss this, let us first consider the action 
\bea
S &=& - \int_M d^2 x \sqrt{-h_2} \, P(R_2) + 2 \int_{\partial M} \,  dt \, \sqrt{-\gamma} \, P'(R_2) \, K \nonumber\\
&& + 
\frac{1}{\sqrt{v_1}} \, 
\int_{\partial M} \, dt \,  \sqrt{-\gamma} \, 
 64 i  (F_{\Upsilon} - {\bar F}_{\Upsilon})   \left( v_2 \, R_2 -2 \right)\;.
\label{bbwr2}
\eea
Note that the boundary term in the last line makes use of the scale $v_1$, introduced in \eqref{constcurv},  
associated with the two-dimensional space-time.
Varying this action with respect to the metric and using \eqref{FFp}
gives rise to the following non-vanishing boundary terms, 
\bea
&&  \int_{\partial M} \sqrt{-\gamma} \; \delta h_{ij} \left(   \, h^{ij}  \nabla_n  P'(R_2) -  n^i \nabla^j  P'(R_2) + n^i \gamma^{jk}   \mathcal{D}_k P'(R_2)
\right)
\nonumber\\
&& - \int_{\partial M}  \sqrt{-\gamma} \, 
  64 i  (F_{\Upsilon} - {\bar F}_{\Upsilon}) \, \left[ v_2  \left( K -  \frac{1}{\sqrt{v_1}} \right) 
   \delta R_2 - \frac{\left(v_2 R_2 -2\right) }{\sqrt{v_1}}\frac{\gamma^{ij}}{2} \delta \gamma_{ij }\right] \;.
\label{btrou}
\eea
Using the decomposition $h^{ij} = \gamma^{ij} + n^i n^j $ and $\gamma^{ij} n_j = 0$, this becomes
\be
 \int_{\partial M} \sqrt{-\gamma} \left\{ \gamma^{ij} \delta \gamma_{ij} \,  \, \nabla_n  P'(R_2) - 64 i  (F_{\Upsilon} - {\bar F}_{\Upsilon}) \, \left[
  v_2  \left( K -  \frac{1}{\sqrt{v_1}} \right) 
 \delta R_2 - \frac{\left(v_2 R_2 -2\right) }{\sqrt{v_1}}\frac{\gamma^{ij}}{2} \delta \gamma_{ij }\right] \right\} \,.
 \label{var3}
\ee
The first two terms can be made to vanish by demanding
\bea
 \nabla_n  P'(R_2) = P''(R_2) \nabla_n R_2 = 0 \;,
  \label{ccur}
 \eea
 and
\bea
K\vert_{\partial M} -  1/\sqrt{v_1} = 0 \;.
\eea
These two conditions are satisfied by constant curvature metrics \eqref{constcurv}.
The third term in \eqref{var3} can be made to vanish by demanding
\bea 
v_2 = v_1 \;.
\label{v1v2}
\eea
Thus, we have set up a well-defined variational principle with Dirichlet boundary conditions that allows for solutions
with $v_2 = v_1$. Solutions with $v_2 \neq v_1$  are also possible in the higher derivative theories that we are considering 
\cite{Cardoso:2006xz},
but we will not be concerned with them in this paper.

Thus, by demanding that the variational principle discussed above allows for solutions with 
constant curvature metrics \eqref{constcurv} satisfying \eqref{v1v2},
we are led to consider the bulk-boundary action \eqref{bbppaar2}, with $S_1$ now 
depending on additional fields and
given by\footnote{Even though $\Upsilon$ is real, we have chosen
to use both $\Upsilon$ and $\bar \Upsilon$ in these expressions to make the symplectic structure manifest.}
\bea
\label{S1fg}
S_1  &=&
2 \int_{\partial M} dt \sqrt{- \gamma} \, \left[ P' (R_2) \,
  K - f(Y, \bar Y, \Upsilon) + 
  \frac{1}{\sqrt{v_1}} 
  32 i  (F_{\Upsilon} - {\bar F}_{\Upsilon})   \left( v_2 \, R_2 -2 \right)
   \right] \nonumber\\
&& +  \tfrac14 \, g_1 (v_1)  \int_{\partial M} dt \,   \sqrt{-\gamma}
 \,  (\pi_I, {\tilde \pi}^I)
\begin{bmatrix}
4 N^{IJ} &  2 N^{I K} R_{KJ} \\
2 R_{IK} N^{KJ} & \;\;
N_{IJ} + R_{IK} N^{KL} R_{LJ}  
\end{bmatrix} \, \begin{pmatrix}
\pi_J \\ 
{\tilde \pi}^J
\end{pmatrix} 
\nonumber\\
&&  + 2   g_2 (v_1)  \int_{\partial M} dt \,   \sqrt{-\gamma} \, 
\left[  
 (Y^I + {\bar Y}^I, F_I + {\bar F}_I )
\begin{pmatrix}
	\pi_I \\ 
	{\tilde \pi}^I
\end{pmatrix} \right. \\
&& \left. \quad 
- 2 i  \Upsilon
\begin{pmatrix}
F_{\Upsilon J} N^{JI} ,  F_{\Upsilon L}  
N^{L K} \bar{F}_{KI} 
\end{pmatrix}
 \begin{pmatrix}
\pi_I \\ 
{\tilde \pi}^I
\end{pmatrix}
+ 2 i \bar{\Upsilon} 
\begin{pmatrix}
 \bar{F}_{\Upsilon J}  N^{JI} , \bar{F}_{\Upsilon L}  
N^{LK} {F}_{KI} 
 \end{pmatrix}  \begin{pmatrix}
\pi_I \\ 
{\tilde \pi}^I
\end{pmatrix}
\right] \;. \nonumber
\eea
The function $f(Y, \bar Y, \Upsilon)$ is taken to be independent of $(\pi^I, {\tilde \pi}_I)$.
The functions $g_1 (v_1)$ and $g_2 (v_1)$ only depend on $v_1$.
Note that $S_1$ is a symplectic scalar: it is constructed out of quantities that transform as scalars, vectors and tensors
under symplectic transformations.  This is a consequence of the underlying electric-magnetic duality of the 
system \cite{deWit:1996gjy,Mohaupt:2011aa,LopesCardoso:2019mlj}.

So far, we discussed the variational principle for the dynamical fields. Let us now turn to the constant scalar fields $v_2, Y^I, \Upsilon$.
Varying the bulk action with respect to these fields gives rise to algebraic relations for these fields. On the other hand, demanding the 
vanishing of the boundary
action $S_1$ under constant changes of $v_2, Y^I, \Upsilon$ results in conditions on the functions $g_1$, $g_2$ and $f$.
These functions are further constrained by demanding that the bulk-boundary action \eqref{bbppaar2}, when evaluated on 
constant curvature solutions to the equations of motion,
is finite. This will be discussed in Subsection \ref{sec:S1deter}. The resulting expressions for  $g_1$, $g_2$ and $f$ will ensure that the 
renormalized on-shell boundary action \eqref{renosba}  describing the coupling of sources is finite, as will be discussed in Subsection \ref{sec:holren}.

\subsection{Equations of motion}   

The equations of motion for the field strengths $e^I$ and $f_J$ that follow from the bulk Lagrangian $H(e^I,f_J)$ given in \eqref{bulkr2lag2},
\bea
\partial_{j} \left( \frac{ \partial  H}{\partial F_{ij}^I} \right)  = 0 \;\;\;,\;\;\; \partial_{j} \left( \frac{ \partial  H}{\partial G_{ij}^I} \right)  = 0 \;,
\eea
  are solved by
  \bea
\frac{ \partial  H}{\partial e^I} = constant = - q_I \;\;\;,\;\;\; \frac{ \partial  H}{\partial f_I} = constant = p^I \;,
\label{Hepq}
\eea
where $(q^I, p_I)$ denote electric and magnetic charges, respectively, in accordance with \eqref{entrofunc}.
Using \eqref{derHef}, this gives
\bea
\label{feqp}
f_I  &=&  \tfrac12 R_{IJ}  e^J + \tfrac12 \frac{\sqrt{-h_2}}{v_2}  \, N_{IJ} \left[ p^J +  2 i \left(Y^J - \bar{Y}^J\right) 
+
4 \Upsilon 
N^{IJ} F_{\Upsilon J}  + 4 \bar{\Upsilon}  N^{IJ} \bar{F}_{\Upsilon J} 
\right] \,, \nonumber\\
e^I &=&  - 2 \frac{\sqrt{-h_2} }{v_2} \, N^{IJ} \left[q_J - \tfrac12 R_{JK} p^K - N_{JK} \left(Y^K + \bar{Y}^K \right)+ 2i \left( \Upsilon F_{\Upsilon J} - 
\bar{ \Upsilon} \bar{F}_{\Upsilon J}\right) 
\right] \,.
\eea
This can be written in terms of symplectic vectors as
\bea
\label{efsympl}
\begin{pmatrix}
f_I \\
- e^I
\end{pmatrix} &=& \frac{ \sqrt{-h_2} }{2 v_2}  \left\{
\begin{bmatrix}
N_{IJ} + R_{IK} N^{KL} R_{LJ}  & \;\;\; - 2 R_{IK} N^{KJ} \\
- 2 N^{I K} R_{KJ} & 4 N^{IJ}
\end{bmatrix} \, 
\begin{pmatrix}
p^J \\ 
q_J
\end{pmatrix}  \right.
 \\
&&  \left.+   
4 
\begin{pmatrix}
F_I + {\bar F}_I  \\
- ( Y^I + {\bar Y}^I ) 
\end{pmatrix}
- 8 i  \Upsilon
\begin{pmatrix}
\bar{F}_{IK} N^{KL} F_{\Upsilon L}  \\
- N^{IJ} F_{\Upsilon J} 
\end{pmatrix}
+ 8 i \bar{\Upsilon} 
\begin{pmatrix}
{F}_{IK} N^{KL} \bar{F}_{\Upsilon L}  \\
- N^{IJ} \bar{F}_{\Upsilon J} 
\end{pmatrix}
\right\} \;. \nonumber
\eea

We now evaluate the associated gauge potentials.
We work in the gauge $A_r^I = \tilde{A}_{I\, r} = 0$.
For
a space-time metric of the form \eqref{metfeff},
 we obtain
 \bea 
  \begin{pmatrix}
\tilde{A}_{t I  }\\
- A_t^I
\end{pmatrix} = \sqrt{v_1} \,  \frac{ 
{\alpha} (t) \, e^{r/\sqrt{v_1}}  }{ {\sqrt{-h_2}} }
\left( 1 - \frac{ \beta}{\alpha} \,  e^{-2r/\sqrt{v_1}} \right)
\begin{pmatrix}
f_I \\
- e^I
\end{pmatrix} + \begin{pmatrix}
\tilde{\mu}_I (t) \\
- \mu^I (t)
\end{pmatrix} \;.
\label{AtAr}
\eea
Here, 
$\tilde{\mu}_I (t) $ and  $\mu^I (t)$ represent $U(1)$ gauge degrees of freedom.
At large $r$,
\bea
\frac{ 
{\alpha} (t) \, e^{r/\sqrt{v_1}}  }{ {\sqrt{-h_2}} }
\left( 1 - \frac{ \beta}{\alpha} \,  e^{-2r/\sqrt{v_1}} \right) \approx \left( 1 - 2  \frac{ \beta}{\alpha} \,  e^{-2r/\sqrt{v_1}} \right) \;.
\label{largeralbet}
\eea
Thus, at 
$r = \infty$, $A^I_t $ and $\tilde{A}_{t I }$ behave as
\be
A^I_t = \sqrt{v_1} \,  e^I  \;, \qquad \tilde{A}_{t I } = \sqrt{v_1} \,  f_I   \;,
\label{Aef}
\ee
and diverge as $e^{r/\sqrt{v_1}}$ in view of \eqref{efsympl}.

Next, we consider the equation of motion for the space-time metric $h_{ij}$  in the presence of $R_2^2$ interactions.
Varying the bulk action with respect to the metric yields, c.f.~\eqref{eq_variation_bulk_F_R},
\bea 
P' (R_2) \, R_{2 \, ij} - \tfrac12 P(R_2) \, h_{ij}\, - h_{ij} \, \Box P' (R_2) + \nabla_i \nabla_j P' (R_2) = T_{ij} \;. 
\label{einsttr}
\eea
In the following we focus on 
constant curvature solutions \eqref{constcurv}. Then, using \eqref{RRh} and \eqref{bulkr2lag2}, and setting $T_{ij} =  \tfrac12 \, T \, h_{ij}$, 
\eqref{einsttr} becomes
\bea
P' (R_2) \, R_2 - P(R_2) \, =  T\;,
\label{traceeins}
\eea
 with 
\bea
T &=& - \frac{1}{v_2} \Big\{
- \tfrac14 \, ( \sqrt{-h_2}/v_2 )^{-2} \, (e^I, f_I)
\begin{bmatrix}
\label{UhY}
N_{IJ} + R_{IK} N^{KL} R_{LJ}  &\;\;\;  - 2 R_{IK} N^{KJ} \\
- 2 N^{I K} R_{KJ} & 4 N^{IJ}
\end{bmatrix} \, 
\begin{pmatrix}
e^J \\ 
f_J
\end{pmatrix}  \nonumber\\
&& \qquad +\left(  \frac{8}{\sqrt{-\Upsilon}} - 2 \right) i (\bar Y^IF_I -Y^I\bar F_I) 
- 2i ( \Upsilon \, F_{\Upsilon} - \bar{\Upsilon} \bar{F}_{\Upsilon} ) 
+ 8 \Upsilon \bar{\Upsilon} \, \bar{F}_{\Upsilon I} N^{IJ} F_{\Upsilon J} 
 \nonumber\\
&&  \qquad \qquad \qquad + 2 \Upsilon F_{\Upsilon I} N^{IJ} (F_J - \bar{F}_{JL} Y^L) + 
 2 \bar{\Upsilon} \bar{F}_{\Upsilon I} N^{IJ} (\bar{F}_J - F_{JL} \bar{Y}^L) 
\nonumber\\
  && \qquad 
  + 2 i (F_\Upsilon-\bar F_\Upsilon) \,
  \left( 32  - 8   \sqrt{-\Upsilon} \right)
  \Big\} 
 \;. 
 \label{T}
\eea
Next, we specialise to constant curvature solutions \eqref{constcurv} that satisfy the condition  \eqref{v1v2},
so that
\bea
v_2 R_2 = 2 \;.
\label{vrc}
\eea
Then, using \eqref{FFp}, the left hand side of~\eqref{traceeins} equals, 
\bea
P' (R_2) \, R_2 - P(R_2) \, = - 16 i (F_\Upsilon-\bar F_\Upsilon) \, v_2 R_2^2 = - 64 i (F_\Upsilon-\bar F_\Upsilon) \,\frac{1 }{v_2} \;.
\eea
Inserting this into \eqref{traceeins} and using \eqref{T} results in 
\bea
&&  64 i (F_\Upsilon-\bar F_\Upsilon) = 
- \tfrac14 \, ( \sqrt{-h_2}/v_2 )^{-2} \, (e^I, f_I)
\begin{bmatrix}
N_{IJ} + R_{IK} N^{KL} R_{LJ}  &\;\;\;  - 2 R_{IK} N^{KJ} \\
- 2 N^{I K} R_{KJ} & 4 N^{IJ}
\end{bmatrix} \, 
\begin{pmatrix}
e^J \\ 
f_J
\end{pmatrix}  \nonumber\\
&& \qquad +\left(  \frac{8}{\sqrt{-\Upsilon}} - 2 \right) i (\bar Y^IF_I -Y^I\bar F_I) 
- 2i ( \Upsilon \, F_{\Upsilon} - \bar{\Upsilon} \bar{F}_{\Upsilon} ) 
+ 8 \Upsilon \bar{\Upsilon} \, \bar{F}_{\Upsilon I} N^{IJ} F_{\Upsilon J} 
 \nonumber\\
&&  \qquad \qquad \qquad + 2 \Upsilon F_{\Upsilon I} N^{IJ} (F_J - \bar{F}_{JL} Y^L) + 
 2 \bar{\Upsilon} \bar{F}_{\Upsilon I} N^{IJ} (\bar{F}_J - F_{JL} \bar{Y}^L) 
\nonumber\\
  && \qquad 
  + 2 i (F_\Upsilon-\bar F_\Upsilon) \,
  \left( 32  - 8   \sqrt{-\Upsilon} \right)
 \;. 
 \label{treins}
\eea
We now turn to a class of solutions with constant curvature metrics \eqref{vrc} that satisfy \eqref{treins}. 
For this class of solutions we have $v_2 = v_1$ in the presence of $R_2^2$ interactions.
We will refer to these solutions
as BPS solutions.

\subsection{Class of solutions}

Large BPS black holes in four space-time dimensions exhibit the attractor mechanism, according to which the scalar fields $Y^I$
supporting the extremal black hole attain specific values at the horizon~\cite{Ferrara:1996dd,Ferrara:1996um}. These attractor values,
which are determined in terms of the electric and magnetic charges carried by the BPS black hole, are obtained by solving 
the attractor equations, which
in the presence of $W^2$ interactions take the form \cite{LopesCardoso:1998tkj}
\bea
Y^I - {\bar Y}^I = i p^I \;\;\;,\;\,\; F_I - {\bar F}_I = i q_I \;\;\;,\;\;\; \Upsilon = - 64 \;,
\label{BPSeq}
\eea
where we recall that the holomorphic function $F(Y, \Upsilon)$ depends on both $Y^I$ and $\Upsilon$, c.f. \eqref{FYUps}.
Using \eqref{relYg2g4}, we infer the BPS relation
\bea
\frac{v_2}{G_4} =  i (\bar Y^IF_I -Y^I\bar F_I )  \;.
\label{v2att}
\eea
Note that the right hand side is dimensionless: it is expressed in terms of the charges through the attractor equations \eqref{BPSeq}.

In the following, we take the constant scalar fields $Y^I$ and $\Upsilon$ to satisfy the attractor equations \eqref{BPSeq}.
Then, substituting the electric/magnetic charges $(q_I, p^I)$ by the expressions \eqref{BPSeq}, and inserting these into \eqref{efsympl}
yields
\bea
\begin{pmatrix}
f_I \\
e^I
\end{pmatrix} &=& \frac{ \sqrt{-h_2} }{v_2}  
\begin{pmatrix}
F_I +{\bar F}_I  \\
Y^I + {\bar Y}^I 
\end{pmatrix} \;,
\label{feY}
\eea
which are indeed the Maxwell field strength expressions for BPS solutions when evaluated in an AdS$_2$ background \cite{Cardoso:2006xz}.

Using the BPS expressions \eqref{feY}, we evaluate
\bea
&& ( \sqrt{-h_2}/v_2 )^{-2} \, (e^I, f_I)
\begin{bmatrix}
\label{UhY}
N_{IJ} + R_{IK} N^{KL} R_{LJ}  & \;\;\; - 2 R_{IK} N^{KJ} \\
- 2 N^{I K} R_{KJ} & 4 N^{IJ}
\end{bmatrix} \, 
\begin{pmatrix}
e^J \\ 
f_J
\end{pmatrix} \nonumber\\
&&= - 4 i (\bar Y^IF_I -Y^I\bar F_I) +  32 \Upsilon \bar{\Upsilon} \, \bar{F}_{\Upsilon I} N^{IJ} F_{\Upsilon J} 
 \nonumber\\
&&  \qquad \qquad \qquad + 8 \Upsilon F_{\Upsilon I} N^{IJ} (F_J - \bar{F}_{JL} Y^L) + 
 8 \bar{\Upsilon} \bar{F}_{\Upsilon I} N^{IJ} (\bar{F}_J - F_{JL} \bar{Y}^L) \;.
 \eea
Then, inserting this into the equation of motion \eqref{treins}, we find that \eqref{treins} is satisfied.

Summarising, 
we will focus on 
solutions to the 
two-dimensional equations of motion
that have a 
constant curvature metric \eqref{metfeff}
satisfying
\eqref{v1v2},
and that are supported by Maxwell field strengths \eqref{feY} and by constant scalar fields that satisfy the BPS attractor equations
\eqref{BPSeq}. These solutions are given by:
\bea
ds^2_2 &=& dr^2 + {h}_{tt} (r,t) \, dt^2  \;\;\;,\;\;\; - {h}_{tt} =\left(  {\alpha} (t) \, e^{r/\sqrt{v_1}} + {\beta} (t) \, e^{-r/\sqrt{v_1}} \right)^2 \;,
 \nonumber\\
  v_1 &=& v_2 
  =  i (\bar Y^IF_I -Y^I\bar F_I ) \, G_4
   \;,  \nonumber\\
Y^I - {\bar Y}^I &=& i p^I \;\;\;,\;\,\; F_I - {\bar F}_I = i q_I \;\;\;,\;\;\; \Upsilon = - 64 \;, \nonumber\\
\begin{pmatrix}
f_I \\
e^I
\end{pmatrix} &=& \frac{ \sqrt{-h_2} }{v_2}  
\begin{pmatrix}
F_I +{\bar F}_I  \\
Y^I + {\bar Y}^I 
\end{pmatrix} \;.
 \label{2dBPS}
\eea
For these solutions, the asymptotic behaviour of $K$ is given by
\bea
K = \frac{1}{\sqrt{v_1}} \left[ 1 - 2\, \frac{\beta}{\alpha} \, e^{-2 r /\sqrt{v_1}} + {\cal O} \left( e^{-4r/\sqrt{v_1}} \right)  \right] \;.
\label{Kasym}
\eea

Regarding the on-shell relations  \eqref{piArenos} for this class of solutions, which are valid at the boundary $\partial M$, we demand that the boundary action $S_1$ in 
\eqref{S1fg} is such that on-shell, 
 \bea
 A_t^{ ren \, I}  \vert_{\partial M}  = \mu^I(t) \;\;\;,\;\;\;  {\tilde A}^{ren}_{I \, t}  \vert_{\partial M } = {\tilde \mu}_I (t) \;.
 \label{atrenos}
 \eea
 This will be discussed in the next subsection.
Then, the on-shell relations \eqref{piArenos} become
 \be
 \label{ospipq}
  \pi_I = - q_I \;, \quad   {\tilde \pi}^I = p^I \;, \quad 
  A_t^{ ren \, I} =  \mu^I(t)  \;, \quad 
   {\tilde A}^{ren}_{I \, t} = {\tilde \mu}_I (t) \;,
  \ee
 and the Dirichlet boundary conditions \eqref{varAren}, 
 when evaluated on the space of solutions to the field
equations, correspond to
\bea
\delta \mu^I (t) \vert_{\partial M} =0 \;\;\;,\;\;\;  \delta {\tilde \mu}_I (t) \vert_{\partial M} =0 \;.
\eea

We note the following two relations which we will use in subsequent sections.
First, using \eqref{feY}, we obtain
\bea
&&   (e^I, f_I) \left[ 2i 
\begin{pmatrix}
F_I - {\bar F}_I \\
- (Y^I - {\bar Y}^I) 
\end{pmatrix}
+ 4 \Upsilon
\begin{pmatrix}
\bar{F}_{IK} N^{KL} F_{\Upsilon L}  \\
- N^{IJ} F_{\Upsilon J} 
\end{pmatrix}
+ 4 \bar{\Upsilon} 
\begin{pmatrix}
{F}_{IK} N^{KL} \bar{F}_{\Upsilon L}  \\
- N^{IJ} \bar{F}_{\Upsilon J} 
\end{pmatrix}
\right] \nonumber\\
&&= \frac{\sqrt{-h_2}}{v_2} \, \left[  4 i \left( {\bar Y}^I F_I - Y^I {\bar F}_I \right)
- 16 \Upsilon \bar{\Upsilon} \, \bar{F}_{\Upsilon I} N^{IJ} F_{\Upsilon J} \right.
 \nonumber\\
&& \left.  \qquad \qquad \qquad -4 \Upsilon F_{\Upsilon I} N^{IJ} (F_J - \bar{F}_{JL} Y^L) -
 4 \bar{\Upsilon} \bar{F}_{\Upsilon I} N^{IJ} (\bar{F}_J - F_{JL} \bar{Y}^L) 
 \right] \;.
 \label{rel1}
\eea
Second, using the on-shell relation \eqref{ospipq} and the attractor equations \eqref{BPSeq}, we obtain
 \bea
 &&  (\pi_I, {\tilde \pi}^I)
\begin{bmatrix}
4 N^{IJ} &  2 N^{I K} R_{KJ} \\
2 R_{IK} N^{KJ} & \;\;\;
N_{IJ} + R_{IK} N^{KL} R_{LJ}  
\end{bmatrix} \, \begin{pmatrix}
\pi_J \\ 
{\tilde \pi}^J
\end{pmatrix} \nonumber\\
&& =  (-q_I, p^I)
\begin{bmatrix}
4 N^{IJ} &  2 N^{I K} R_{KJ} \\
2 R_{IK} N^{KJ} & \;\;\;
N_{IJ} + R_{IK} N^{KL} R_{LJ}  
\end{bmatrix} \, \begin{pmatrix}
-q_J \\ 
p^J
\end{pmatrix} \nonumber\\
&&=  (p^I, q_I)
\begin{bmatrix}
N_{IJ} + R_{IK} N^{KL} R_{LJ} & \;\;\; - 2 R_{IK} N^{KJ} \\
 2 N^{I K} R_{KJ} & 4 N^{IJ} \\
\end{bmatrix} \, \begin{pmatrix}
p^J\\
q_J
\end{pmatrix} \nonumber\\
&& = - 4 i  (\bar Y^IF_I -Y^I\bar F_I) + 32 \Upsilon \bar{\Upsilon} \, \bar{F}_{\Upsilon I} N^{IJ} F_{\Upsilon J} 
 \nonumber\\
&& \qquad    -8 \Upsilon F_{\Upsilon I} N^{IJ} (F_J - \bar{F}_{JL} Y^L) -
 8 \bar{\Upsilon} \bar{F}_{\Upsilon I} N^{IJ} (\bar{F}_J - F_{JL} \bar{Y}^L) 
\;.
\label{rel2}
\eea

Next,  we determine the functions $f, g_1, g_2$ 
that appear in the boundary action
$S_1$, c.f. \eqref{S1fg}.

\subsection{Determining $S_1$ \label{sec:S1deter}}

To determine the functions $f, g_1, g_2$ in $S_1$ given in \eqref{S1fg}, we proceed as follows.  
First we demand that at the boundary $\partial M$, which in the coordinate system \eqref{metfeff}  is located at $r = \infty$, the on-shell 
relation \eqref{atrenos} is satisfied.  Next, we demand that the bulk-boundary action \eqref{bbppaar2} is finite when evaluated
on a  solution given by \eqref{2dBPS}. These two conditions lead to restrictions on the form of the functions $f, g_1, g_2$ for the case of solutions of the form  \eqref{2dBPS}.
Finally, 
we demand that at $r= \infty$, $S_1$ vanishes under constant variations of $v_2, Y^I, \Upsilon$.

To impose the relation  \eqref{atrenos}, we consider \eqref{piArenos}.
For simplicity we 
 demand that on-shell,
 \bea
 g_1 (v_1)  = g_2 (v_1)  \equiv g ( v_1) \;.
 \eea
Then, using the on-shell relation \eqref{piArenos} as well as \eqref{efsympl} with $\sqrt{-h_2}\vert_{\partial M} = \sqrt{-\gamma}$,
we obtain 
\bea
&& \frac{\delta S_1}{\delta \pi_K} = g (v_1)   \,  \sqrt{-\gamma}
\,  \left( 2 \,  \pi_J N^{JK} + {\tilde \pi}^I R_{IJ} N^{JK} +  2 (Y + {\bar Y})^K  - 4 i \Upsilon \, F_{\Upsilon J} \, N^{JK} 
+ 4 i \bar{\Upsilon} 
\bar{F}_{\Upsilon J}  N^{JK} 
\right)  \nonumber\\
&&= 
g (v_1)   \,  \sqrt{-\gamma}
\,  \left( 2 \,  \frac{\partial H}{\partial e^J} \,  N^{JK} + \frac{\partial H}{\partial f_I} \, R_{IJ} N^{JK} +  2 (Y + {\bar Y})^K 
- 4 i \Upsilon \, F_{\Upsilon J} \, N^{JK} + 4 i \bar{\Upsilon} 
\bar{F}_{\Upsilon J}  N^{JK} \right) \nonumber\\
&& =
g (v_1) \, v_2 \,  \,   e^K \;. 
\eea 
  Now we demand that on-shell,
 \bea
 g (v_1) \, v_2  = \sqrt{v_1} \;.
 \label{valueg}
 \eea
 Then, using \eqref{Aef}, we obtain
\bea
A^{ren \, I}_t \vert_{r = \infty}  = \left( A^I_t - \frac{\delta S_1}{\delta \pi_I} \right) \vert_{r = \infty} = 
\left( A_t^I -  \sqrt{v_1}  \,   e^I  \right) \vert_{r = \infty}= \mu^I (t) \;,
\eea
and similarly, 
\bea
 {\tilde A}_{I \, t}^{ren} \vert_{r = \infty}= {\tilde \mu}_I (t) \;.
\eea

Next, we impose finiteness of the bulk-boundary action \eqref{bbppaar2} when evaluated on  solutions of the form  \eqref{2dBPS}, i.e. we demand
that all the terms that diverge at the boundary $\partial M$ cancel out. Evaluating the various combinations in \eqref{bbppaar2} at $r = \infty$ 
using \eqref{rel1} and \eqref{rel2} as well as 
$K\vert_{r = \infty} = 1/\sqrt{v_1}$ and
$\sqrt{-\Upsilon} = 8$, we obtain the following on-shell expressions,
\bea
\int dr H(e,f) = \frac{ \sqrt{v_1}}{v_2}  \, \sqrt{-\gamma} \, \Big[
i (\bar Y^IF_I -Y^I\bar F_I) 
-128 i (F_{\Upsilon} - \bar{F}_{\Upsilon} ) 
 \Big] \;,
 \label{Hf}
\eea
\bea
2 \int_{\partial M} dt \sqrt{- \gamma} \, \left[ P' (R_2)  \,
  K - f(Y, \bar Y, \Upsilon) \right] =   2\int_{\partial M} dt \sqrt{- \gamma} \left[ \frac{P' (R_2)}{\sqrt{v_1}} 
        -  f(Y, \bar Y, \Upsilon) 
    \right] \;,
    \label{KF}
\eea
\bea
 - \int dt \left(\pi_I  \; \,, \; \tilde{\pi}^I \right)
\, \begin{pmatrix}
	A^I_t  - A^{ren \, I}_t  \\
	{\tilde A}_{I \, t}   - {\tilde A}^{ren}_{I \, t} 
\end{pmatrix}  &=& \sqrt{v_1}  \int_{\partial M} dt \,  \left( q_I \,  e^I  - p^I f_I \right)  \nonumber\\
&=&  - 2  \frac{\sqrt{v_1} }{v_2}  \int_{\partial M} dt \,\sqrt{- \gamma} \, i  (\bar Y^IF_I -Y^I\bar F_I) \;,
 \label{AArneY}
\eea
\bea
&&\tfrac14  g_1( v_1)  \int_{\partial M} dt \,  \sqrt{-\gamma}
 \,  (\pi_I, {\tilde \pi}^I)
\begin{bmatrix}
4 N^{IJ} &  2 N^{I K} R_{KJ} \\
2 R_{IK} N^{KJ} & \;\;\;
N_{IJ} + R_{IK} N^{KL} R_{LJ}  
\end{bmatrix} \, \begin{pmatrix}
\pi_J \\ 
{\tilde \pi}^J
\end{pmatrix} =  \nonumber\\
&&  g_1( v_1)  \int_{\partial M} dt \,  \sqrt{-\gamma} 
\Big[ -  i  (\bar Y^IF_I -Y^I\bar F_I) + 8 \Upsilon \bar{\Upsilon} \, \bar{F}_{\Upsilon I} N^{IJ} F_{\Upsilon J} 
 \nonumber\\
&& \qquad \qquad \qquad    -2 \Upsilon F_{\Upsilon I} N^{IJ} (F_J - \bar{F}_{JL} Y^L) -
 2 \bar{\Upsilon} \bar{F}_{\Upsilon I} N^{IJ} (\bar{F}_J - F_{JL} \bar{Y}^L) \Big]
\;, \nonumber\\
&& 2   g_2( v_1)  \int_{\partial M} dt \,   \sqrt{-\gamma}\, 
 \left(Y^I + \bar{ Y}^I, F_I + {\bar F}_I \right)
 \begin{pmatrix}
\pi_I \\ 
{\tilde \pi}^I
\end{pmatrix} 
= g_2(  v_1) \int_{\partial M} dt \,  \sqrt{-\gamma} \; 4i\left( \bar Y^IF_I -Y^I\bar F_I \right) \;, \nonumber\\
&&2   g_2(  v_1)  \int_{\partial M} dt \,   \sqrt{-\gamma}\, 
\Big[- 2 i  \Upsilon
\begin{pmatrix}
F_{\Upsilon J} N^{JI} ,  F_{\Upsilon L}  
N^{L K} \bar{F}_{KI} 
\end{pmatrix}
 \begin{pmatrix}
\pi_I \\ 
{\tilde \pi}^I
\end{pmatrix} + c.c. \Big]
= \nonumber\\
&&
g_2( v_1) \int_{\partial M} dt \,  \sqrt{-\gamma} \,  \Big[ 4 \Upsilon \, F_{\Upsilon J} N^{JI}  (F_I - {\bar F}_{IK} Y^K ) 
- 8 \Upsilon {\bar \Upsilon} \, F_{\Upsilon J} N^{JI}   {\bar F}_{\Upsilon I} + c.c.  \Big] \;.
\label{ppig22}
\eea
Adding up \eqref{Hf}, the term proportional to $K$ in \eqref{KF}, and \eqref{AArneY} gives zero. Then, 
adding up all the remaining combinations and imposing $v_1 = v_2$ and $g_1 = g_2 = g = 1/\sqrt{v_1} $,
we obtain for the bulk-boundary action at $r = \infty$, 
\bea
\label{stotinf}
S_{total} &=& 
 \int_{\partial M} dt \,  \sqrt{-\gamma} \Big\{ \frac{1}{\sqrt{v_1}} 
 \Big[  3 i (\bar Y^IF_I -Y^I\bar F_I ) - 8 \Upsilon {\bar \Upsilon} \, F_{\Upsilon J} N^{JI}   {\bar F}_{\Upsilon I} \nonumber\\
&& + 2 \Upsilon \, F_{\Upsilon J} N^{JI}  (F_I - {\bar F}_{IK} Y^K ) 
+  2 {\bar \Upsilon} \, {\bar F}_{\Upsilon J} N^{JI}  ({\bar F}_I - F_{IK} {\bar Y}^K ) 
 \Big]  \nonumber\\
&&- 2 f(Y, \bar Y, \Upsilon) 
\Big\} \;.
\eea
Since $\sqrt{-\gamma}$ diverges at $r = \infty$, we demand that the terms in the big bracket in \eqref{stotinf} vanish. 
This yields the on-shell value for $2f$,
\bea \label{fbpsonshellr2}
2 f(Y, \bar Y, \Upsilon)  &=& 
\frac{1}{\sqrt{v_1}} 
\Big[  3 i (\bar Y^IF_I -Y^I\bar F_I )  - 8 \Upsilon {\bar \Upsilon} \, F_{\Upsilon J} N^{JI}   {\bar F}_{\Upsilon I}  \\
&& + 2 \Upsilon \, F_{\Upsilon J} N^{JI}  (F_I - {\bar F}_{IK} Y^K ) 
+  2 {\bar \Upsilon} \, {\bar F}_{\Upsilon J} N^{JI}  ({\bar F}_I - F_{IK} {\bar Y}^K ) 
\Big] \; .\nonumber
\eea
We stress that this holds for solutions of the form  \eqref{2dBPS}.

Next, we discuss the variation of $S_1$ with respect to the constant fields  $v_2, Y^I, \Upsilon$. For simplicity,
we take the functions $g_1$ and $g_2$ in $S_1$ to equal their on-shell value \eqref{valueg},
\bea
g_1 = g_2 = \frac{1}{\sqrt{v_1}} \;.
\label{g1g2v}
\eea
Then, varying $S_1$ with respect to $v_2, Y^I, \Upsilon$ for this choice of functions $g_1$ and $g_2$, and demanding
the vanishing of these variations at $r=\infty$ results in the following expression for the function $2f$,
\bea
2f (Y, \bar Y, \Upsilon) &=& \frac{1}{\sqrt{v_1}} \Big[
  i ({\bar Y}^I F_I - Y^I {\bar F}_I ) \left( 2 + \frac{8}{\sqrt{-\Upsilon}} \right) 
 - 8 \Upsilon {\bar \Upsilon} \, F_{\Upsilon J} N^{JI}   {\bar F}_{\Upsilon I}        \nonumber\\
&& \qquad +  2 \Upsilon \, F_{\Upsilon J} N^{JI}  (F_I - {\bar F}_{IK} Y^K ) 
+  2 {\bar \Upsilon} \, {\bar F}_{\Upsilon J} N^{JI}  ({\bar F}_I - F_{IK} {\bar Y}^K )   \nonumber\\
&& \qquad  + 16 i \left( F_{\Upsilon} - {\bar F}_{\Upsilon} \right) \left( \sqrt{- \Upsilon} - 8 \right) \Big] \;,
\label{value2f}
\eea
where we have imposed the on-shell condition \eqref{fbpsonshellr2}. Note that all the terms in \eqref{value2f}
are symplectic functions. Thus, for the choice of functions \eqref{g1g2v}, this is the unique expression for $2f$ that 
satisfies  \eqref{fbpsonshellr2}.  In the absence of $R_2^2$ interactions, i.e. when $F_{\Upsilon} =0$, the expression for
$2f$ simplifies and becomes
\bea
2f (Y, \bar Y, \Upsilon) &=& \frac{1}{\sqrt{v_1}}\,
  i ({\bar Y}^I F_I - Y^I {\bar F}_I ) \left( 2 + \frac{8}{\sqrt{-\Upsilon}} \right) \;.
  \label{2fwr}
  \eea
  Using \eqref{2fwr}, it is 
  straightforward to verify that the variation of $S_1$ with respect to $v_2, Y^I, \Upsilon$ indeed vanishes at $r=\infty$.
Here one uses that on-shell, $\partial_I V_{BH} = 0$, where we defined
\bea
V_{BH} = 
   (\pi_I, {\tilde \pi}^I)
\begin{bmatrix}
4 N^{IJ} &  2 N^{I K} R_{KJ} \\
2 R_{IK} N^{KJ} & \;\;\;
N_{IJ} + R_{IK} N^{KL} R_{LJ}  
\end{bmatrix} \, \begin{pmatrix}
\pi_J \\ 
{\tilde \pi}^J
\end{pmatrix} \;.
\eea
 In the presence of $R_2^2$ interactions, the calculation is more involved and makes
repeated use of the special geometry identities \eqref{sgrel} as well as of the on-shell relations
\bea
\partial_I V_{BH}  \vert_{\rm on-shell} &=& 4i F_{IJK} Y^J Y^K - 4i \Sigma_K \, N^{KL} F_{I L P} Y^P 
+ i \Sigma_J \, N^{JP} \, F_{I PQ} \, N^{QK} \Sigma_K \;, \nonumber\\
\partial_{\Upsilon}V_{BH}  \vert_{\rm on-shell} &=& 4i F_{\Upsilon JK} Y^J Y^K - 4i \Sigma_K \, N^{KL} F_{\Upsilon L P} Y^P 
+ i \Sigma_J \, N^{JP} \, F_{\Upsilon PQ} \, N^{QK} \Sigma_K \;, \nonumber\\
\eea
where here we treat $\Upsilon$ as  a holomorphic variable, and where
\bea
\Sigma_I = 4 i \left( \Upsilon \, F_{\Upsilon I} - {\bar \Upsilon} \, {\bar F}_{\Upsilon I} \right) \;.
\eea

Thus, we have determined the boundary action $S_1$: taking $g_1$ and $g_2$ to be given by \eqref{g1g2v}, the function $2f$
is uniquely specified by \eqref{value2f}. The resulting expression for $S_1$ takes the form
\bea
\label{s1g1g2fsub}
S_1  &=&
 \int_{\partial M} dt \sqrt{- \gamma} \, \Big\{2 \, P' (R_2) \,
  K -  \frac{1}{\sqrt{v_1}} \Big[
  i ({\bar Y}^I F_I - Y^I {\bar F}_I ) \left( 2 + \frac{8}{\sqrt{-\Upsilon}} \right) 
 - 8 \Upsilon {\bar \Upsilon} \, F_{\Upsilon J} N^{JI}   {\bar F}_{\Upsilon I}        \nonumber\\
&& \qquad +  2 \Upsilon \, F_{\Upsilon J} N^{JI}  (F_I - {\bar F}_{IK} Y^K ) 
+  2 {\bar \Upsilon} \, {\bar F}_{\Upsilon J} N^{JI}  ({\bar F}_I - F_{IK} {\bar Y}^K )   
\nonumber\\
&& \qquad 
 + 16 i \left( F_{\Upsilon} - {\bar F}_{\Upsilon} \right) \left( \sqrt{- \Upsilon} - 4 v_2 \, R_2 \right) \Big] 
   \Big\} \nonumber\\
&& +  \frac{1}{4 \, \sqrt{v_1}}  \int_{\partial M} dt \,   \sqrt{-\gamma}
 \,  (\pi_I, {\tilde \pi}^I)
\begin{bmatrix}
4 N^{IJ} &  2 N^{I K} R_{KJ} \\
2 R_{IK} N^{KJ} & 
N_{IJ} + R_{IK} N^{KL} R_{LJ}  
\end{bmatrix} \, \begin{pmatrix}
\pi_J \\ 
{\tilde \pi}^J
\end{pmatrix} 
\nonumber\\
&&  + \frac{2}{ \sqrt{v_1}}  \int_{\partial M} dt \,   \sqrt{-\gamma} \, 
\left[  
 (Y^I + {\bar Y}^I, F_I + {\bar F}_I )
\begin{pmatrix}
	\pi_I \\ 
	{\tilde \pi}^I
\end{pmatrix} \right. \\
&& \left. \quad 
- 2 i  \Upsilon
\begin{pmatrix}
F_{\Upsilon J} N^{JI} ,  F_{\Upsilon L}  
N^{L K} \bar{F}_{KI} 
\end{pmatrix}
 \begin{pmatrix}
\pi_I \\ 
{\tilde \pi}^I
\end{pmatrix}
+ 2 i \bar{\Upsilon} 
\begin{pmatrix}
 \bar{F}_{\Upsilon J}  N^{JI} , \bar{F}_{\Upsilon L}  
N^{LK} {F}_{KI} 
 \end{pmatrix}  \begin{pmatrix}
\pi_I \\ 
{\tilde \pi}^I
\end{pmatrix}
\right] \;. \nonumber
\eea
Once again, we emphasize that this holds for solutions of the form  \eqref{2dBPS}.

\subsection{Holographic renormalization keeping $G_4$ fixed \label{sec:holren} }

To formulate the consistent variational principle discussed in Subsection  \ref{sec:var-prin-r2}, we considered
arbitrary variations of the dynamical fields and subjected them to the boundary conditions detailed there.  In this subsection, 
we investigate the behaviour under specific variations, namely 
perturbations around BPS solutions of the form \eqref{2dBPS}.
We do so while keeping the scalar field combination in \eqref{G4} fixed, which is equated to $G_4^{-1}$.

We recall from \eqref{metfeff} and \eqref{ospipq}  that the asymptotic (large $r$) behaviour of the bulk metric and of $A_t^{ren I}, {\tilde A}_ {I t}^{ren}$
is
\be
h_{tt} = - \alpha^2 (t) \, e^{2 r /\sqrt{v_1}} \;\;\;,\;\;\;  A_t^{ren \, I} = \mu^I(t) \;\;\;,\;\;\;  {\tilde A}_ {I t}^{ren} = {\tilde \mu}_I (t) \;,
\ee
which results in the following asymptotic variations in the space of solutions,
\be
\delta \gamma_{tt} =  e^{2 r /\sqrt{v_1}} \, \delta \left(-  \alpha^2 \right)  \;\;\;,\;\;\;  \delta A_t^{ren \, I} = \delta \mu^I  \;\;\;,\;\;\;  
\delta {\tilde A}_ {I t}^{ren} = \delta {\tilde \mu}_I  \;.
\ee
Following \cite{Cvetic:2016eiv}, we supplement these asymptotic variations with a perturbation of the scalar $v_2$, as follows. While so far we took $v_2$
to be a constant scalar, we now consider its asymptotic perturbation 
\be
\delta v_2 =  e^{ r /\sqrt{v_1}} \, \delta \nu  \;.
\label{pertv2}
\ee
Here, $\nu(t)$ acts a source for an irrelevant operator of scaling dimension $2$ in the boundary
theory c.f. \eqref{renacvar}, and hence, $\delta v_2$ describes a perturbation around two-dimensional BPS solutions of the form \eqref{2dBPS}.
We do not fluctuate the scalar fields $X^I$ and $\hat A$, and therefore we stay in the Poincar\'e gauge given in 
\eqref{G4}.
Using the relations \eqref{rescalXA}, we infer that the perturbation of $v_2$ induces the following perturbation of $Y^I$ and $\Upsilon$,
\bea
\delta Y^I = Y^I \, \frac{\delta v_2}{v_2} \;\;\;,\;\;\; \delta \Upsilon = 2 \, \Upsilon \, \frac{\delta v_2}{v_2} \;.
\label{varYupsv}
\eea
Under these asymptotic variations, the bulk-boundary action \eqref{bbppaar2} will transform into
\bea
\label{renacvar}
\delta S_{total} = \int_{\partial M} dt  \left( \pi^{tt} \, \delta \gamma_{tt} + \pi_{v_2} \, \delta v_2 + \pi_I \, \delta A_t^{ren \, I} + 
{\tilde \pi}^I \, \delta {\tilde A} _{I t}^{ren } 
\right) \;, 
\eea
where the conjugate momenta $ \pi^{tt},  \pi_{v_2}, \pi_I , {\tilde \pi}^I$ are evaluated on-shell. 
Using $S_1$ given in \eqref{s1g1g2fsub}, we will show below that 
the variation of the action \eqref{renacvar}
is finite.
One then refers to \eqref{renacvar} as the variation of the 
renormalized on-shell action. It can be written in terms of sources $\alpha^2, \nu, \mu^I,  {\tilde \mu}_I$ as
\bea
\label{renacvar2}
\delta S_{total} = \int_{\partial M} dt \, \alpha  \left[ \hat{\pi}^{tt} \, \delta \left(-\alpha^2 \right)+ \hat{\pi}_{v_2} \, \delta \nu + \hat{\pi}_I \, \delta \mu^I + 
\hat{\tilde \pi}^I \, \delta {\tilde \mu} _{I }
\right]  \;, 
\eea
where the finite 1-point functions conjugate to the sources $\alpha^2$, $\nu$, $\mu_{I }$ and $\tilde{\mu}_I$ are, respectively \cite{Skenderis:2002wp},
\bea
\hat{\pi}^{t}_t &=&  -  \alpha^2 \hat{\pi}^{tt}  =  \lim_{r \rightarrow \infty} \left( -  \alpha^2 \, \frac{e^{2r/\sqrt{v_1}}}{\alpha} \, \pi^{tt} \right) 
=   \lim_{r \rightarrow \infty } \left(  e^{  r /\sqrt{v_1}}  \frac{ \gamma_{tt} }{ \sqrt{ - \gamma} } \frac{\delta S_{total}}{\delta  \gamma_{tt}} \right) 
 \;, \nonumber\\
\hat{\pi}_{v_2} &=&  \lim_{r \rightarrow \infty} \left( \frac{e^{ r/\sqrt{v_1}} }{\alpha} \,  \pi_{v_2} \right) = 
\lim_{r \rightarrow \infty}  \left(   \frac{ e^{  2r /\sqrt{v_1}}  }{\sqrt{ - \gamma} } \frac{\delta S_{total}}{\delta v_2} \right) \;,
\nonumber\\
\hat{\pi}_I &=& \frac{\pi_I}{\alpha}  \;\;\;,\;\;\;  \hat{\tilde \pi}^I  = \frac{{\tilde \pi}^I}{\alpha} \;.
\label{1pfunc}
\eea
This conforms with the rule for computing 1-point functions of dual operators $O$ in the boundary CFT$_d$: given a supergravity field
 with asymptotic expansion 
\bea
{\cal F} (r, y) =  e^{-2 m r /\sqrt{v_1}}  \left( f_{(0)} (y) + \dots \right) \;,
\eea
its associated 1-point function is \cite{Skenderis:2002wp} 
\bea
\langle O \rangle
=  \lim_{r \rightarrow \infty}  \left(  \frac{e^{ (d- 2m) r /\sqrt{v_1}} }{\sqrt{ - \gamma} } \,  \frac{\delta S_{total}}{\delta  {\cal F}} \right)   \;.
\label{Ograv}
\eea
In our case we have $d=1$ and for  ${\cal F} = \gamma_{tt}$ we have $m=-1$, while for ${\cal F} = v_2$ we have $m = - 1/2$.
The renormalized on-shell action, defined by
\bea
S_{ren} [ f_{(0)} ] =   \lim_{r \rightarrow \infty} S_{total} [  {\cal F}] \;, 
\label{renosba}
\eea
describes the coupling of the sources  $f_{(0)}$ to the 1-point functions $\langle O \rangle $ \cite{Skenderis:2002wp}.

Now we compute the 1-point functions \eqref{1pfunc}. We begin by computing $\pi^{tt}$.
Varying the bulk-boundary action \eqref{bbppaar2} with respect to the metric, and taking into account that only
the terms $\int d^2 x H + S_1$ will vary, we obtain, using \eqref{S1fg} with $g_1 = g_2 = 1/\sqrt{v_1}$,
\bea
\pi^{tt} &=& \frac12 \sqrt{-\gamma} \,  \gamma^{tt} \left\{ -2 f(Y, \bar Y, \Upsilon)
+\frac{1}{4 \, \sqrt{v_1}} 
 \,  (\pi_I, {\tilde \pi}^I)
\begin{bmatrix}
4 N^{IJ} &  2 N^{I K} R_{KJ} \\
2 R_{IK} N^{KJ} & \;\;\;
N_{IJ} + R_{IK} N^{KL} R_{LJ}  
\end{bmatrix} \, \begin{pmatrix}
\pi_J \\ 
{\tilde \pi}^J
\end{pmatrix} \right. \nonumber\\
&& 
\qquad \qquad 
 + \frac{2}{  \sqrt{v_1}} \, 
 \left[  (Y^I + {\bar Y}^I, F_I + {\bar F}_I )
 \begin{pmatrix}
\pi_I \\ 
{\tilde \pi}^I
\end{pmatrix}
- 2 i  \Upsilon
\begin{pmatrix}
F_{\Upsilon J} N^{JI} ,  F_{\Upsilon L}  
N^{L K} \bar{F}_{KI} 
\end{pmatrix}
 \begin{pmatrix}
\pi_I \\ 
{\tilde \pi}^I
\end{pmatrix} \right. 
\nonumber\\
&& \left. \left.  \qquad \qquad \qquad \qquad + 2 i \bar{\Upsilon} 
\begin{pmatrix}
 \bar{F}_{\Upsilon J}  N^{JI} , \bar{F}_{\Upsilon L}  
N^{LK} {F}_{KI} 
 \end{pmatrix}  \begin{pmatrix}
\pi_I \\ 
{\tilde \pi}^I
\end{pmatrix}
\right]
\right\}\;.
\eea
We evaluate this expression on solutions of the form \eqref{2dBPS}.
Using the on-shell relations \eqref{rel2}, 
we obtain 
the 
on-shell value
\bea
\label{pttgrav}
\pi^{tt} &=& \frac12 \sqrt{-\gamma}  \,  \gamma^{tt} 
 \left\{ -2 f(Y, \bar Y,\Upsilon)
+  \frac{1}{\sqrt{v_1}} 
 \Big[  3 i (\bar Y^IF_I -Y^I\bar F_I ) \right.  \\
&& \left. + 2 \Upsilon \, F_{\Upsilon J} N^{JI}  (F_I - {\bar F}_{IK} Y^K ) 
+  2 {\bar \Upsilon} \, {\bar F}_{\Upsilon J} N^{JI}  ({\bar F}_I - F_{IK} {\bar Y}^K ) 
- 8 \Upsilon {\bar \Upsilon} \, F_{\Upsilon J} N^{JI}   {\bar F}_{\Upsilon I} \Big]  \right\} \;. \nonumber
\eea
Using \eqref{fbpsonshellr2}, we find that the terms in the big bracket vanish and hence, on-shell, $\pi^{tt} $ 
vanishes identically,
\bea
\hat{\pi}^{t}_t   = 0 \;.
\eea
Since higher $n$-point functions are obtained by multiple differentiation of $S_{total}$ with respect
to $\gamma_{tt}$ (c.f. \eqref{Ograv}), and since these expressions are obtained by varying \eqref{pttgrav}
with respect to $\gamma_{tt}$, it follows that also the associated $2$-point function vanishes, and hence the central charge
of the putative boundary CFT$_1$ vanishes.

Next, we compute $\pi_{v_2}$.
We vary the bulk-boundary action \eqref{bbppaar2} with respect to $v_2$,
keeping in mind that the variation 
\eqref{pertv2} induces the variation of the scalar fields $Y^I$ and $\Upsilon$ given in \eqref{varYupsv}.
Only the boundary action $S_1$, given in \eqref{s1g1g2fsub},
contributes.  At large $r$, the divergent terms in $\pi_{v_2}$ all cancel out, since
we established in Subsection \ref{sec:S1deter} that the variation of $S_1$ with respect to the scalars $v_2, Y^I, \Upsilon$ 
vanishes at $r = \infty$. The subleading contribution, however, is non-vanishing. It stems from the term proportional to $K$
in \eqref{s1g1g2fsub} and is given by
\bea
\pi_{v_2}  = -2 \alpha \, e^{- r /\sqrt{v_1}} 
\, i (\bar Y^IF_I -Y^I\bar F_I )
\frac{1}{\sqrt{v_1} v_2} \, \frac{\beta}{\alpha}  + {\cal O} \left(e^{- 3r /\sqrt{v_1}}  \right) \;.
\eea
Thus, when evaluated on a solution of the form \eqref{2dBPS}, we obtain the asymptotic on-shell value
\bea
{\hat  \pi} _{v_2} =  -2 \, i (\bar Y^IF_I -Y^I\bar F_I )
    \frac{1}{\sqrt{v_1} v_2} \,\frac{ \beta }{\alpha}=  -
    \frac{2}{\sqrt{v_1} } \, \frac{\beta}{\alpha \, G_4} 
    \;,
\eea
where we used \eqref{v2att} and \eqref{1pfunc}.
Note that ${\hat  \pi} _{v_2} $ has length dimension $-3$, in accordance with \eqref{renacvar2}.

Summarizing, we obtain for the 1-point functions \eqref{1pfunc},
\bea
\hat{\pi}^{t}_t  
&=& 0 \;, \nonumber\\
{\hat  \pi} _{v_2} &=&  -    \frac{2}{\sqrt{v_1} } \, \frac{ \beta }{\alpha \, G_4}  \;, \nonumber\\
\hat{\pi}_I &=& - \frac{q_I}{\alpha}  \;\;\;,\;\;\;  \hat{\tilde \pi}^I  = \frac{p^I}{\alpha} \;,
\label{1ppi}
\eea
and for the renormalized on-shell action \eqref{renosba},
\bea
S_{ren} = \int_{\partial M} dt  \left( - \frac{2}{\sqrt{v_1}} \, \frac{\beta}{G_4}  \, \nu  - q_I \, \mu ^{I} + 
p^I \, {\tilde \mu} _I + {\cal O} (\nu^2) 
\right) \;.
\label{ronsac}
\eea
The mode $\beta (t)$ does not affect the asymptotic behaviour of the metric given in \eqref{2dBPS}. Nevertheless, it 
enters in the action \eqref{ronsac} through its coupling to the source $\nu$, and hence 
the mode $\beta (t)$ corresponds to the one-point function of a scalar operator 
 of scaling dimension $2$ in the boundary theory, as pointed out in \cite{Cvetic:2016eiv}.

\section{Penrose-Brown-Henneaux diffeomorphisms and asymptotic symmetries}\label{sec_PBH_analysis}

In this section we discuss the anomalous transformation behaviour of ${\hat  \pi} _{v_2}$ under asymptotic symmetries,
following the analysis given in \cite{Papadimitriou:2005ii,Cvetic:2016eiv,Castro2018ffi}, which we first summarize.

We begin by reviewing the residual transformations that leave the Fefferman-Graham gauge of the metric~\eqref{metfeff} and the radial gauge 
\eqref{delAr} of the gauge fields invariant. 
The residual diffeomorphisms in 
these transformations are known as Penrose-Brown-Henneaux (PBH) diffeomorphisms~\cite{Brown:1986nw,Penrose1986ca,Imbimbo:1999bj} which, due to the presence of gauge fields in the
present context, have to be supplemented by $U(1)$ gauge transformations~\cite{Papadimitriou:2005ii,Cvetic:2016eiv,Castro2018ffi}.
Next, we focus on asymptotic symmetries, to obtain the transformation rules for the holographic dual operators that were computed in the previous section. 
Asymptotic symmetries are PBH transformations that that leave the sources invariant.

The starting point 
consists in considering an infinitesimal bulk diffeomorphism $\xi^i$ and 
gauge transformations $\Lambda^I, \tilde{\Lambda}_I$, so that the fields transform as
\be
\delta_\xi h_{ij}  = \mathcal{L}_\xi h_{ij} \,, \quad \delta_{\xi + \Lambda^I} A^I_i =  \mathcal{L}_\xi A^I_i + \partial_i \Lambda^I \,, \quad \delta_{\xi + \tilde{\Lambda}_I} \tilde{A}_{I\, i} =  \mathcal{L}_\xi \tilde{A}_{I\,i} + \partial_i \tilde{\Lambda}_I \,, \quad \delta_{\xi} v_2 = \mathcal{L}_\xi v_2 \,.
\ee 
Preservation of the Fefferman-Graham gauge and of the radial gauge of $A_i$ requires
\be \label{eq_PBH_diff_eqs}
\mathcal{L}_\xi h_{rr}= \mathcal{L}_\xi h_{rt} =0 \,, \quad \mathcal{L}_\xi A^I_r + \partial_r \Lambda^I =0 \,, \quad \mathcal{L}_\xi \tilde{A}_{I\,r} + \partial_r \tilde{\Lambda}_I =0  \,. 
\ee
Solving these equations for $\xi^t$, $\xi^r$ yields 
\be \label{eq_PBH_vector_field}
\xi^t = \varepsilon (t) + \partial_t \sigma (t) \int_r^{\infty} h^{tt} (r', t) \, dr'  \,, \quad \xi^r = \sigma (t) \;,
\ee
with $\varepsilon(t)$ and $\sigma(t)$ arbitrary functions of the time-like coordinate $t$. Note that $\varepsilon$ generates boundary diffeomorphisms, 
while $\sigma$ generates boundary Weyl transformations. Moreover, solving~\eqref{eq_PBH_diff_eqs} for the gauge transformations $\Lambda^I$ and $\tilde{\Lambda}_I$ we obtain
\bea
\Lambda^I &=& \varphi^I (t) - \partial_t \sigma (t) \, \int_r^{\infty}  h^{tt}(r', t)  \, A_t^I (r', t)  \, dr' \,,  \nn \\
{\tilde \Lambda}_I &=& {\tilde \varphi}_I (t) - \partial_t \sigma (t) \, \int_r^{\infty}  h^{tt}(r', t)  \, {\tilde A}_{I t} (r', t)  \, dr' \,, \label{eq_PBH_Lambdas}
\eea
with $\varphi^I(t)$ and $\tilde{\varphi}_I(t)$ arbitrary functions of the time-like coordinate $t$ which generate the boundary gauge transformations. 

Plugging~\eqref{eq_PBH_vector_field} and~\eqref{eq_PBH_Lambdas} into the metric in the gauge~\eqref{metfeff}, the gauge fields using~\eqref{AtAr} and $v_2$ using~\eqref{pertv2} and taking the large $r$ limit, we find
\bea \label{diff}
 \delta_\mathrm{PBH}\, \alpha &=& \frac{\sigma}{\sqrt{v_1}} \,  \, \alpha  + \partial_t ( \varepsilon \,   \alpha) \;, \nonumber\\
\delta_\mathrm{PBH} \, \beta &=&  \partial_t \left(\varepsilon \, \beta \right)- \frac{\sigma}{\sqrt{v_1}} \, \beta 
 - \frac{\sqrt{v_1}}{2} \, \partial_t \left( \frac{\partial_t \sigma}{\alpha} \right) \;, \nn \\
\delta_\mathrm{PBH}\,  \mu^I &=&  \partial_t ( \varepsilon \, \mu^I + \varphi^I) \;, \nn \\
\delta_\mathrm{PBH} \, {\tilde \mu}_I &=&  \partial_t ( \varepsilon \, {\tilde \mu}_I + {\tilde \varphi}_I) \;, \nn\\
\delta_\mathrm{PBH} \, \nu &=& \varepsilon \, \partial_t \nu + \frac{\sigma}{\sqrt{v_1}} \nu \,.
 \label{infdiffe}
 \eea
Additionally, from~\eqref{infdiffe} we find that the PBH transformations act on the one-point functions \eqref{1ppi} as
\bea
\delta_\mathrm{PBH} \, \hat{\pi}_{v_2} &=& \varepsilon \, \partial_t \hat{\pi}_{v_2} - 2 \frac{\sigma}{\sqrt{v_1}} \hat{\pi}_{v_2} + \frac{1 }{\alpha \, G_4}  \partial_t\left( \frac{\partial_t \sigma}{\alpha}  \right) \,, \nn  \\ 
\delta_\mathrm{PBH} \, \hat{\pi}_I  &=&  - \left(  \frac{\sigma}{\sqrt{v_1}}  +\frac{\partial_t \left(\varepsilon\, \alpha\right)}{\alpha}  \right)  \hat{\pi}_I   \,, \nn \\ 
\delta_\mathrm{PBH} \, \hat{\tilde{\pi}}^I  &=& - \left(  \frac{\sigma}{\sqrt{v_1}}  +\frac{\partial_t \left(\varepsilon\, \alpha\right)}{\alpha}  \right)  \hat{\tilde{\pi}}^I  \,. \label{eq_1_point_PBH_transf}
\eea
Finally, we compute the change of the renormalized on-shell action \eqref{ronsac}
under these transformations. 
Using~\eqref{infdiffe}, we find that under PBH transformations
this action transforms as
\be
\delta_\mathrm{PBH} \, S_\mathrm{ren} = \frac{1}{G_4} \int_{\partial M}  dt \;  \sigma\,  \partial_t\left(\frac{\partial_t \nu}{\alpha}\right) \,,
\ee
where we have dropped total derivative terms. Hence, in the presence of a source for 
$v_2$, the boundary theory ceases 
being invariant under Weyl transformations. On the other hand, it is always invariant under time reparameterizations and $U(1)$ gauge transformations.

We now focus on asymptotic symmetries, i.e. on the symmetry transformations \eqref{infdiffe} that leave the sources invariant.\footnote{ 
Hence, asymptotic symmetry
transformations depend crucially on the choice of sources in the boundary theory \cite{Cvetic:2016eiv}.}
Since we are interested in the case of constant $v_2$, following \eqref{pertv2},
we focus strictly on the symmetries that exist when the source $\nu$ is set to zero \cite{Cvetic:2016eiv}. Therefore, we wish to consider the PBH transformations for which  
\be
\delta \alpha =  \delta {\mu}^I = \delta  \tilde{\mu}_I = 0 \;.
\label{asympt}
\ee
The set of functions $\varepsilon,\sigma,\varphi^I,\tilde{\varphi}_I$ which fulfil these conditions is given by
\be
\varepsilon= \frac{\zeta(t) }{\alpha} \,, \quad \sigma = - \sqrt{v_1} \, \frac{ \partial_t \left( \varepsilon \, \alpha  \right) }{\alpha} \,,  \quad \varphi^I = - \varepsilon \, \mu^I + k^I \,, \quad \tilde{\varphi}_I = - \varepsilon \, \tilde{\mu}_I +\tilde{ k}_I \,, \label{eq_BKV}
\ee
with $\zeta(t)$ an arbitrary function of time and $k^I, \tilde{ k}_I$ arbitrary constants. Since we have imposed $\delta \alpha =0$, we conclude from~\eqref{eq_1_point_PBH_transf} that $\hat{\pi}_I$ and $\hat{\tilde{\pi}}^I$ are invariant under the subset of symmetries given in~\eqref{eq_BKV}. On the other hand, $\hat{\pi}_{v_2}$ transforms non-trivially as
\be
\delta_{sym} \hat{\pi}_{v_2}  = \zeta\, \partial_+ \hat{\pi}_{v_2} +2 \partial_+ \zeta \, \hat{\pi}_{v_2} -  
\lambda\,
 \partial_+^3 \zeta \,,
\label{delpiv2}
\ee
where we have defined $dx^+ = \alpha dt$, and where
\bea
\lambda = \frac{\sqrt{v_1} }{  G_4}  =  \frac{\sqrt{v_1}}{v_2} \, i \left( \bar{ Y}^I F_I - Y^I \bar{ F}_I   \right)  \;. 
\label{cads2}
\eea
Note that $\lambda$ has length dimension $-1$, as it must, since $\hat{\pi}_{v_2}$ has length dimensions $-3$.
Now we recall the relation (c.f. \eqref{2dBPS})
\bea
v_1 = v_2 = \left( p^I F_I - q_I Y^I \right) G_4 = \left( p^I {\bar F} _I - q_I {\bar Y} ^I \right) G_4 = \frac{v_2 \, G_4}{2 \sqrt{- h_2}}   \left( p^I f_I - q_I e^I \right) 
\;,
\eea
which we use to 
bring $\lambda$ into the suggestive form,
\bea
\lambda = \frac{\sqrt{v_1}  }{ 2 \sqrt{- h_2} }  \ \left( p^I f_I - q_I e^I \right) \;.
\eea
Using \eqref{efsympl}, this behaves schematically as
\bea
\lambda \sim \frac{Q^2}{\sqrt{v_1}}  \;,
\eea
where $Q$ denotes a charge, and where the two-dimensional scale $\sqrt{v_1}$ has length dimension $1$.
Hence, in units of $\sqrt{v_1}$, $\lambda$ behaves as $\lambda \sim Q^2$.

Thus, \eqref{delpiv2} shows that $\hat{\pi}_{v_2} $ is the 1-point function of a scalar operator of conformal dimension $2$, transforming anomalously under asymptotic symmetries.
 Under the infinitesimal transformation
$\zeta (x^+) = \epsilon_n  \, (x^+)^{n+1} $
with constant $\epsilon_n, \, n \in \mathbb{Z}$, we get
\bea
\delta_{sym} \hat{\pi}_{v_2}    &=& -  \epsilon_n  \, L_n \,  \hat{\pi}_{v_2} - \lambda \, (n^2 -1) n \,  \epsilon_n \,  (x^+)^{n-2} \;,
\eea
where $L_n$ denotes the differential operator
\bea \label{eq_def_mode_Ln}
l_n = - (x^+)^{n+1} \, \partial_+    - \Delta \,  (n+1)   \, (x^+)^n  
  \;\;\;,\;\;\; \Delta = 2 \;.
\eea
The $L_n$ furnish a representation of the Witt algebra
\bea
[L_n, L_m] = (n-m) \, L_{n + m} \;.
\label{1ptwitt}
\eea
The $sl(2, \mathbb{R}) $ subalgebra of the Witt algebra is generated by $L_{-1} = -H, \, L_0 = - D, \, L_1 = - K$, where
\bea
H = \partial_+ \;\;\;,\;\;\; D = x^+ \, \partial_+ + \Delta  \;\;\;,\;\;\; K = \left( x^+ \right)^2  \, \partial_+ + 2 \Delta \, x^+ \;.
\eea
Now we set $\hat{\pi}^{t}_t  = \langle T \rangle$ and
 $ \hat{\pi}_{v_2} = \langle O \rangle$, and we perform
 the mode expansion 
\bea
T = \sum_{n \in \mathbb{Z}} {\hat L}_n \, (x^+)^{-n-2} \;\;\;,\;\;\; O = \sum_{m \in \mathbb{Z}} {\hat J}_m \, (x^+)^{-m-\Delta} \;,
\eea
where here ${\hat L}_n, {\hat J}_m$ denote operators that act on states in the Hilbert space of the putative CFT$_1$.
Then, defining $ \delta \hat{\pi}_{v_2} = \langle \delta O \rangle$ and choosing
 $ \delta_{\epsilon_n} O = [{\hat L}_n, O]$,
we obtain 
\bea
[{\hat L}_n, {\hat J}_m] = \left( n (\Delta -1) -m \right) {\hat J}_{n + m} - \lambda (n^2 -1) n \, \delta_{n+m,2 - \Delta} \;.
\label{ljalg}
\eea
Then, inserting $\Delta = 2$, gives
\bea
[{\hat L}_n, {\hat J}_m] = \left( n -m \right) {\hat J}_{n + m} - \lambda \, n \,  (n^2 -1)  \, \delta_{n+m,0} \;.
\label{ljalgdel}
\eea
Beyond the generic features of the algebra recounted above, details of how these operators act on the Hilbert space will require more specific knowledge of the CFT in question.

\section{AdS$_2$ holography and Wald entropy \label{sec:AdS2Wald}}

We now propose to identify how Wald entropy is encoded holographically. In order to do so, we observe that the Wald entropy 
of four-dimensional black holes scales quadratically in the charges. Using the attractor equations given in 
\eqref{BPSeq} as well as the homogeneity relation \eqref{FYUps}, 
we are led to consider a perturbation whose effect is to induce simultaneously a scaling of order $1$ in 
$Y^I$ and of order $2$ in $\Upsilon$.
Namely, we consider the following asymptotic perturbation around BPS solutions of the form \eqref{2dBPS}.
\bea
\delta = e^{r/\sqrt{v_1}} \, \delta \Omega \, {\cal D} \;,
\eea
where ${\cal D}$ denotes the vector field 
\bea
{\cal D} = Y^I \, \frac{\partial}{\partial Y^I} + 2 \Upsilon \, \frac{\partial}{\partial \Upsilon} + c.c.
\label{vecfD}
\eea
along which the perturbation is taken, and where $\Omega (t)$ denotes a source of length dimension zero.
Using \eqref{rescalXA}, this perturbation can be viewed as being induced by the change
\bea
\delta X^I = \tfrac12 X^I \, \delta \Omega \;\;\;,\;\;\; \delta {\hat A} =  {\hat A} \,  \delta \Omega \;\;\;,\;\;\; \delta w = \tfrac12 w \, \delta \Omega \;,
\eea
which, using \eqref{G4}, induces a change of $ i \left( {\bar X}^I F_I (X, \hat{A}) - X^I {\bar F}_I ({\bar X}, \bar{\hat{A}} )\right)$, 
\bea
\delta \left[  i \left( {\bar X}^I F_I (X, \hat{A}) - X^I {\bar F}_I ({\bar X}, \bar{\hat{A}} )\right) 
\right]=   i \left( {\bar X}^I F_I (X, \hat{A}) - X^I {\bar F}_I ({\bar X}, \bar{\hat{A}} )\right) \,  \delta \Omega \;.
\eea
Note that this perturbation does not induce a variation of $v_2$, which is in accordance with \eqref{relYg2g4}.

Under this asymptotic variation, the bulk-boundary action \eqref{bbppaar2} will transform into
\bea
\delta S_{total} = \int_{\partial M} dt  \,e^{r/\sqrt{v_1}} \left( Y^I \Pi_I  + 2 \Upsilon \, \Pi_{\Upsilon} + c.c. \right) 
\delta \Omega \;,
\eea
where the conjugate momenta $ \Pi_I$ and  $\Pi_{\Upsilon}$, defined by
\bea
\Pi_I = \frac{\delta S_{total}}{\delta Y^I} \;\;\;,\;\;\; \Pi_{\Upsilon} = \frac{\delta  S_{total}}{\delta \Upsilon} \;,
\eea
are evaluated on-shell. We again demand this variation to be finite, thereby obtaining the
renormalized on-shell variation of the action, 
\bea
\delta S_{total} = \int_{\partial M} dt \, \alpha  \, \hat{\Pi} \, \delta \Omega \;,
\eea
with
\bea
\hat{\Pi} &=& Y^I \hat{\Pi}_I  + 2 \Upsilon \, \hat{\Pi}_{\Upsilon} + c.c. , \nonumber\\
\hat{\Pi}_{I} &=&  \lim_{r \rightarrow \infty} \left( \frac{e^{ r/\sqrt{v_1}} }{\alpha} \,  \Pi_{I} \right) = 
\lim_{r \rightarrow \infty}  \left(   \frac{ e^{  2r /\sqrt{v_1}}  }{\sqrt{ - \gamma} } \frac{\delta S_{total}}{\delta Y^I} \right) \;,
\nonumber\\
\hat{\Pi}_{\Upsilon} &=&  \lim_{r \rightarrow \infty} \left( \frac{e^{ r/\sqrt{v_1}} }{\alpha} \,  \Pi_{\Upsilon} \right) = 
\lim_{r \rightarrow \infty}  \left(   \frac{ e^{  2r /\sqrt{v_1}}  }{\sqrt{ - \gamma} } \frac{\delta S_{total}}{\delta \Upsilon} \right) \;.
\eea
Note that $\hat{\Pi}$ has length dimension $-1$.

Next, we note the following relation, which holds when evaluated on a solution \eqref{2dBPS},
\bea
{\cal D} P'(R_2)  =  P'(R_2) \;,
\eea
with $P'(R_2)$ given in \eqref{FFp}.
Then, proceeding as in the case of the computation of $\hat{\pi}_{v_2}$, we obtain, using \eqref{Kasym},
\bea
\hat{\Pi} = - \frac{4}{\sqrt{v_1}} \, P'(R_2) \, \frac{\beta}{\alpha} \;.
\eea

Under the asymptotic symmetry transformation \eqref{eq_BKV}, 
the one-point function $\hat{\Pi}$, evaluated on a solution \eqref{2dBPS} 
with $\alpha = 1, \beta =0$, displays the anomalous behaviour,
\bea
\delta_{sym} \hat{\Pi} \vert_{anomalous} = -2 P'(R_2) \, \sqrt{v_1} \, \partial_t^3 \varepsilon  \;,
\label{anomvarg4}
\eea
Therefore, the coefficient  $P'(R_2)$, which equals the Wald entropy of BPS black holes when evaluated on a 
solution \eqref{2dBPS}, c.f. \eqref{bpswalde}, is
in fact encoded in the anomalous transformation of an operator $O$ on the boundary, whose one-point function 
$\hat{\Pi }= \langle O \rangle$ is obtained by the holographic prescription of varying the action with respect to the 
corresponding bulk field.  Viewing the bulk-boundary action as a function of $z^A= Y^A/Y^0, \, \Upsilon/(Y^0)^2,$ and $\Upsilon (Y^0)^2$,
the first two combinations are invariant under the action of the 
vector field ${\cal D}$ given in \eqref{vecfD}, while the third combination is not.  We may thus identify the corresponding bulk
field with  $\ln \left[ (Y^0)^2 \, \Upsilon \right] $.
Therefore, the anomaly in the asymptotic symmetry transformation of the CFT$_1$ operator holographically dual to this bulk field
generates Wald's Noether charge \eqref{bpswalde} in two dimensions, 
and hence the corresponding BPS black hole entropy in four dimensions.

\section{Dimensional reduction keeping $\kappa_2^2$ fixed: the 4d/5d connection
\label{sec:4d5d}} 

In the above, we discussed the reduction of four-dimensional 
$\mathcal{N}=2$ supergravity theories with Weyl squared interactions to two dimensions on a product space-time with line element \eqref{backgr},
where all physical quantities were measured in units of $G_4$. 
Now we discuss the reduced theory in units of  $1/\kappa_2^2 = B^2/G_4$, c.f. \eqref{G2G4}. 
This discussion is relevant for the 4d/5d connection \cite{Gaiotto:2005gf,Castro:2007sd,Banerjee:2011ts}.
Below we will discuss the interpretation of $B^2$.
Recall that in two dimensions, Newton's constant $\kappa_2^2$ has length dimension zero. 

To this end, we perform the coordinate transformation
\bea
r = e^{-\psi/2} \, \tilde{r} \;\;\;,\;\;\; t= e^{-\psi/2}  \, \tilde{t}  \;,
\eea
and the redefinition
\bea
v_1 = e^{-\psi}  \, \tilde{ v}_1 \;,
\eea
so that the two-dimensional metric becomes
\bea
ds_2^2 =  h_{ij} \, dx^i dx^j =  e^{-\psi}  {d\tilde{s}_2}^2  = e^{-\psi} \, \tilde{h}_{ij} \, dx^i dx^j \;\;\;,\;\; x^i = (\tilde{r}, \tilde{t}) \;,
\eea
with ${d\tilde{s}_2}^2$ given in Fefferman-Graham form by
\bea
{d\tilde{s}_2}^2 &=& d {\tilde r}^2 + {\tilde h}_{{\tilde t}  {\tilde t} } ({\tilde r} , {\tilde t}) \, d {\tilde t}^2  \;.
\eea
The constant curvature condition $v_1 \, R_2 = 2$ becomes ${\tilde v}_1 {\tilde R}_2 = 2$, in which case $\tilde{h}_{ij} $
takes the form
\be \label{eq_tilde_htt}
\tilde{h}_{tt} = - \left( \alpha\left( \tilde{t} \right)  e^{\tilde{r}/\sqrt{\tilde{v}_1}}  + \beta\left(\tilde{t} \right) e^{-\tilde{r}/\sqrt{\tilde{v}_1}   } \right)^2\,.
\ee

We note the relations
\bea
\sqrt{- h_2} = e^{-\psi} \, \sqrt{- {\tilde h}_2} \;\;\;,\;\;\;  R_2 = e^{\psi} \, {\tilde R}_2 \;\;\;,\;\;\;
\sqrt{-\gamma} = e^{-\psi/2}\,  \sqrt{- \tilde{ \gamma}} \;\;\;,\;\;\; K= e^{\psi/2} \tilde{K} \;,
\eea
and hence, $\sqrt{-\gamma} \, K = \sqrt{- \tilde{ \gamma}}  \, \tilde{K} , \; \sqrt{-\gamma} / \sqrt{v_1} = 
\sqrt{- \tilde{ \gamma}} /  \sqrt{{\tilde v}_1} $. These are the combinations that appear in the boundary action $S_1$ given in \eqref{s1g1g2fsub}.
Having expressed $S_1$ in terms of $\tilde{v}_1$, we now view $\tilde{v}_1$ as a fixed scale.

In addition, 
we substitute $v_2$ by
\bea
v_2 = e^{-\psi} B^2 
\label{v2B}
\eea
in the bulk-boundary action \eqref{bbppaar2},
and subsequently convert $G_4$ into $\kappa_2^2$. Note that the substitution \eqref{v2B} induces a redefinition of the constant fields $Y^I$ and
$\Upsilon$. Namely, using \eqref{rescalXA}, we obtain
\bea
Y^I = e^{- \psi} \; \tilde{Y}^I \;\;\;,\;\;\, \Upsilon = e^{-2\psi} \; \tilde{\Upsilon} \;,
\eea
where
\begin{eqnarray}
   \tilde{Y}^I = \ft14 B^2\, {\bar w} \, X^I \,,\quad
 \tilde{ \Upsilon} = \ft{1}{16} B^4 \, {\bar w}^2 \, {\hat A} = - \ft14
  B^4\, \vert w \vert^4  \;.
  \label{rescalXAB}
  \end{eqnarray}
Then, we regard the bulk-boundary action $S_{total}$ as a function of ${\tilde h}_{ij}, e^{-\psi}$, the constant scalar fields $\tilde{Y}^I, \tilde{\Upsilon}$
and the gauge fields, at fixed parameters $B^2$ and ${\tilde v}_1$. We note that the on-shell condition $v_1 = v_2$ becomes $ {\tilde v}_1 = B^2$.
The rewriting of the bulk-boundary action  \eqref{bbppaar2} in terms of these rescaled fields is performed by using the following relations.

The term $ \sqrt{-h_2} \, P(R_2)$ in the bulk action \eqref{bulkr2lag2}, 
with $P$ given in  \eqref{FFp}, becomes 
\bea
\sqrt{-h_2} P(R_2) = \sqrt{- {\tilde h}_2  } \, e^{-\psi}  P\left(e^\psi \tilde{R}_2   \right) = \sqrt{-\tilde{h}_2}  \, \tilde{P} \left(  \tilde{R}_2  \right) \;,
\eea
with
\bea
\tilde{P} \left(\tilde{R}_2 \right) &=&  
\frac{e^{-\psi}}{2 \kappa_2^2}
 \, \tilde{R}_2
  - 2 i (F_{\tilde \Upsilon} -\bar F_{\tilde \Upsilon}) \,
  \Big ( 8\, B^2   \, \tilde{R}^2_2 - 32 \, \tilde{R}_2 - 4  \, e^{-\psi}\, \tilde{R}_2 \,  \sqrt{- \tilde{\Upsilon}} \Big) \;, \nonumber\\
  \tilde{P}' \left(\tilde{R}_2 \right) &=&   \frac{d \tilde{P}}{d \tilde{R}_2 }=  \frac{e^{-\psi}}{2 \kappa_2^2}
 - 2 i (F_{\tilde \Upsilon} -\bar F_{\tilde \Upsilon}) \,
  \Big ( 16\, B^2  \, \tilde{R}_2 - 32   - 4  \, e^{-\psi}\, \sqrt{- \tilde{\Upsilon}} \Big)\;.
  \label{PPtR}
\eea
It follows that
\bea
\frac{\partial    \,  \tilde{P}' \left(\tilde{R}_2 \right)}{\partial \psi} = - \frac{e^{-\psi} }{ 2 \kappa_2^2}   -8 i (F_{\tilde \Upsilon}-\bar F_{\tilde \Upsilon}) \,
\Big ( \, e^{-\psi}\, \sqrt{- \tilde{\Upsilon}} \Big)  \;.
\label{kappsi}
\eea
The term proportional to $K$ in the boundary action $S_1$ given in \eqref{s1g1g2fsub} becomes
\bea
2 \int_{\partial M} d\tilde{t} \sqrt{-\tilde{ \gamma}}  \,  \tilde{P}' \left(\tilde{ R}_2\right)\, \tilde{K} \;.
\eea
The variation of the boundary action $S_1$ with respect to ${\tilde Y}^I, {\tilde \Upsilon}$ vanishes, as a consequence of
having established in Subsection \ref{sec:S1deter} that the variation of the boundary action $S_1$ with respect to ${Y}^I, {\Upsilon}$ vanishes.

Henceforth, we omit the tildes on all the fields, for notational simplicity.
The bulk-boundary action $S_{total}$ will then contain the terms
\bea
&&- \int_M d^2 x \sqrt{-h_2}  \, {\tilde P}(R_2)+ 2 \int_{\partial M} d {\tilde t}  \sqrt{-\gamma}   \, {\tilde  P}'(R_2) \, K \nonumber\\
&& =  \frac{1}{2\kappa_2^2} \left[ - \int_M d^2 x \sqrt{-h_2} \, e^{-\psi} \, \left( R_2 + \dots \right)  + 2 \int_{\partial M} d {\tilde t} \sqrt{-\gamma} \, e^{-\psi}  \, 
\left( K + \dots \right) 
\right] \;.
\eea
where the ellipsis refers to the remaining terms in \eqref{PPtR}.

Now consider computing the on-shell value of the momentum conjugate to $\psi$, which we denote by
$\hat{\pi}_{\psi}$,
\bea
\hat{\pi}_{\psi} &=&  \lim_{{\tilde r} \rightarrow \infty} \left( \frac{e^{ {\tilde r}/\sqrt{v_1} }}{{\alpha}} \,  \pi_{\psi} \right) = 
\lim_{{\tilde r} \rightarrow \infty}  \left(   \frac{ e^{  2 {\tilde r} / \sqrt{v_1}}  }{\sqrt{ - {\gamma}} } \frac{\delta S_{total}}{\delta \psi} \right) \;.
\eea
Using \eqref{kappsi} and \eqref{Kasym}, we obtain
\bea
\hat{\pi}_{\psi} =  \frac{2 e^{-\psi }}{\sqrt{v_1}} \frac{  \beta}{\alpha} \left[ \frac{1}{\kappa_2^2}   + 16 i \left(F_\Upsilon - \bar{ F}_\Upsilon\right) \sqrt{-\Upsilon}  \right]\;,
\eea
where we recall that on-shell $\sqrt{v_1} =B$, and since $B$ has length dimension $1$, $\hat{\pi}_{\psi} $ has length dimension $-1$.

While so far we took $\psi$ to be a constant scalar, we now consider its perturbation,
of the form
\bea
 \delta \psi = e^{{\tilde r}/\sqrt{v_1}} \, \delta \Xi \;,
 \eea
 where $\Xi (t)$ 
  acts a source for an irrelevant operator of scaling dimension $2$ in the boundary
theory. The above perturbation is mutually exclusive to the one considered in 
\eqref{pertv2}. In the case considered in \eqref{pertv2}, the field combination $ i \left( {\bar X}^I F_I (X, \hat{A}) - X^I {\bar F}_I ({\bar X}, \bar{\hat{A}} )\right)$
given in \eqref{G4} does not fluctuate, while in the case considered here, it is the ratio \eqref{G2G4} that is kept fixed.
Then, recall from \eqref{renacvar2} that the renormalized on-shell action contains a term of the form
\bea
S_{ren} = \int_{\partial M} d {\tilde t} \, \alpha \left( \hat{\pi}_{\psi}  \, \Xi + \dots \right) \;,
\label{srenpsi}
\eea
where the ellipsis refers to additional terms that are not important for the present discussion.
To compare with  \cite{Cvetic:2016eiv}, we introduce
\bea \label{eq_tilde_alpha_beta}
  \tilde{\alpha} = \alpha \;\;\;, \;\;\;   \tilde{\beta}  = e^{-\psi} \beta \;.
\eea
Hence, in terms of $\tilde{\alpha}$ and $ \tilde{\beta}$, \eqref{srenpsi} becomes
\bea
S_{ren} = \int_{\partial M} d {\tilde t} \, \tilde{\alpha}  \left( {\hat{\pi}}_{\psi}  \, \Xi + \dots \right) \;,
\eea
with 
\bea
 {\hat{\pi}}_{\psi}  
 = \frac{2  }{B} \frac{  \tilde{\beta}}{\tilde{\alpha}} \left[ \frac{1}{\kappa_2^2}   + 16 i \left(F_\Upsilon - \bar{ F}_\Upsilon\right) \sqrt{-\Upsilon}  \right]\;,
 \eea
in agreement with \cite{Cvetic:2016eiv} upon setting $\tilde{L} = B$ in their expression.  From~\eqref{eq_tilde_alpha_beta} we find that $\tilde{ \alpha}$ transforms as in \eqref{infdiffe} while the transformation of $\tilde{ \beta}$ is modified into \cite{Cvetic:2016eiv} 
\be \label{eq_variation_tilde_beta_Cvetic}
\delta_{\mathrm{PBH}} \tilde{ \beta} = \partial_{\tilde t} \left(\varepsilon \tilde{\beta} \right) - \frac{\sigma}{ \sqrt{v_1}} \tilde{\beta} - \frac{\sqrt{v_1}}{2} e^{-\psi } \partial_{\tilde t} \left( \frac{\partial_{\tilde t} \sigma}{\tilde{\alpha}} \right) \,. 
\ee
Then, defining $ {\cal O}_{\psi} = - {\hat{\pi}}_{\psi} $, and using the results \eqref{infdiffe},
\eqref{eq_BKV} and~\eqref{eq_variation_tilde_beta_Cvetic} for the variation of $\tilde{\beta}/\tilde{\alpha}$ under asymptotic symmetries,
we obtain
\be
\delta_{sym} {\cal O}_{\psi}  =   \tilde{ \zeta} \, \partial_+ \mathcal{O}_\psi +2 \partial_+ \tilde{ \zeta} \, \mathcal{O}_\psi     - 4 \,  B \, e^{-3 \psi}   \left[ \frac{1}{\kappa_2^2}   + 16 i \left(F_{\tilde \Upsilon} - \bar{ F}_{\tilde \Upsilon} \right) \sqrt{- \tilde \Upsilon}  \right] \,
\partial_+^3 {\tilde \zeta} \;,
\label{opsizet}
\ee
where here we have reinstated the tildes on the rescaled fields, for clarity, and 
where on the right hand side we used the definitions of \cite{Cvetic:2016eiv} ,
\bea
\partial_+ = \frac{e^{\psi}}{2 {\tilde \alpha}} \, \partial_{\tilde t} \;\;\;,\;\;\; {\tilde \zeta} ({\tilde t}) = 2\, \varepsilon ({\tilde t}) \, {\tilde \alpha} ({\tilde t}) \,  e^{-\psi} \;.
\eea
Thus, we have established that the anomalous variation of  ${\hat{\pi}}_{\psi} $ under asymptotic symmetry variations
gets modified by $R^2_2$ corrections. The anomalous variation is proportional to the Wald entropy~\cite{Wald:1993nt}
of the associated four-dimensional
BPS black holes,
\be
\delta_{sym} {\cal O}_{\psi} \vert_{anomalous} =  - 4 \,  B \, e^{-2 \psi} \,   \frac{{\cal S}_{\rm Wald}}{\pi }
\, 
\partial_+^3 {\tilde \zeta} \;,
\label{anomOsym}
\ee
with the Wald entropy given by \cite{LopesCardoso:1998tkj} (c.f. Appendix \ref{sec:Wald_entropy} )
\bea
{\cal S}_{\rm Wald}/\pi =  \, 2 P' (R_2) \vert_{BPS} =   i (\bar Y^IF_I -Y^I\bar F_I) 
+128 i (F_{\Upsilon} - \bar{F}_{\Upsilon} )   \:,
\eea
with the right hand side expressed in terms of charges through the attractor equations given in \eqref{2dBPS}.

Now we return to the redefinition \eqref{v2B} and we discuss the meaning of $B^2$ by using the 4d/5d connection.
To lift the discussion to five dimensions, we restrict to 
four-dimensional BPS solutions that are supported by a subset of charges given by $(q_0, p^A)$ (with $A = 1, \dots, n$).
These solutions can be lifted to solutions of five-dimensional $\mathcal{N}=2$ supergravity theories in the presence of $R^2$ 
interactions~\cite{Gaiotto:2005gf,Castro:2007sd,Banerjee:2011ts}.
These five-dimensional theories can, in turn, be reduced to three dimensions, where one can use the 
AdS$_3$/CFT$_2$ correspondence to further study these solutions.  Thus, solutions supported by charges $(q_0, p^A)$ can be 
analysed both from the point of view
of the AdS$_2$/CFT$_1$ correspondence as well as from the point of view of the AdS$_3$/CFT$_2$ correspondence.

When viewed from a five-dimensional point of view,  the associated five-dimensional line element is
\bea
ds_5^2 &=& e^{\psi} \, ds_4^2 + e^{- 2 \psi} (dx^5 -  A^0)^2 \nonumber\\
&=&  e^{\psi} \,  \left( ds_2^2 + v_2 \, d \Omega_2^2 \right)+ e^{- 2 \psi} (dx^5 - A^0)^2 \nonumber\\
&=&  e^{\psi} \,  ds_2^2 + B^2 \, d \Omega_2^2 + e^{- 2 \psi} (dx^5 - A^0)^2 \nonumber\\
& =&  d\tilde{s}_2^2 + B^2 \, d \Omega_2^2 + e^{- 2 \psi} (dx^5 - A^0)^2 \;,
\label{line542}
\eea
where we used \eqref{backgr}, and where
$B^2 = v_2 \, e^{\psi} $.

To lift the discussion to five dimensions, 
we now consider the four-dimensional $\mathcal{N}=2$ supergravity theory based on
\bea
F(Y, \Upsilon) = - \tfrac16 \, \frac{C_{ABC} Y^A Y^B Y^C }{Y^0} - \tfrac{1}{24} \,  \tfrac{1}{64} \, c_{2A} \, \frac{Y^A}{Y^0} \, \Upsilon \;,
\eea
where $C_{ABC}$ and $c_{2A}$ denote constants associated with Calabi-Yau threefold data. Following \cite{LopesCardoso:1998tkj}, we take
$q_0 < 0$ and $p^A > 0$. Solving the attractor equations given in \eqref{2dBPS} for the $Y^I$ results in  \cite{LopesCardoso:1998tkj}
\bea
\frac{1}{Y^0} = \frac12 \, \sqrt{\frac{|q_0|}{p^3_L} } \;\;\;,\;\;\; Y^A = \frac{i}{2} \, p^A \;,
\eea
where we defined \cite{Castro:2007sd}
\bea
p^3_L = \frac16 \left( C_{ABC} p^A p^B p^C + c_{2A} p^A  \right) > 0 \; ,\; p^3_R = \frac16 \left( C_{ABC} p^A p^B p^C + \frac12 \, c_{2A} p^A  \right) > 0 \;.
\eea
For $v_2$ one obtains
\bea
\frac{v_2}{G_4} = p^A F_A - q_0 Y^0 = \frac{p^3_R }{Y^0} \;,
\eea
while $e^{-\psi}$ is given by \cite{Castro:2007sd}
\bea
e^{- \psi} = p_R  \sqrt{\frac{|q_0|}{p^3_L} } \;.
\eea
It follows that $B^2$, $1/ {\kappa_2^2 }$ and $e^{-\psi}$ can be expressed as
\bea
B^2 = \frac12 \, G_4 \, p^2_R \;\;\;,\;\;\;
\frac{1}{\kappa_2^2 }= \frac12 \, p^2_R \;\;\;,\;\;\; e^{-\psi} = \frac{2 \sqrt{2}}{ Y^0 \, \kappa_2} \;.
\label{kap2pR}
\eea
The BPS entropy is given by
\bea
{\cal S}_{\rm Wald} = 2 \pi \sqrt{ | q_0|  p^3_L } \;.
\eea
Now consider the combination 
\bea
 B \, e^{\psi} \,  \frac{{\cal S}_{\rm Wald} }{\pi} = \sqrt{2} \sqrt{G_4} \, p_L^3 \;.
 \eea
Assuming the validity of the relation (4.24c)\footnote{This was explicitly demonstrated in \cite{Cvetic:2016eiv} at the 2-derivative level, and here we assume 
that it continues to hold at the $R^2$ corrected level.} in \cite{Cvetic:2016eiv}, we obtain for the $++$ component of the 2d CFT stress tensor, 
\bea
\kappa_3^2 \, \tau_{++} =  \kappa_2^2 \, \frac{e^{3 \psi}}{4} \,  {\cal O}_{\psi} + \dots \;,
\label{451}\eea
where the ellipses refers to non-anomalous terms that are not important for the present discussion.
Using \eqref{anomOsym}, we find that $\tau_{++} $
transforms anomalously, 
\bea
 \delta \tau_{++} = -  \frac{ \kappa_2^2}{\kappa_3^2}  \, B \, e^{ \psi} \,   \frac{{\cal S}_{\rm Wald}}{\pi } \, \partial_+^3 \zeta 
= -  \frac{ \kappa_2^2}{\kappa_3^2} \, \sqrt{2} \sqrt{G_4} \, p_L^3 \, \partial_+^3 \zeta   = - \frac{c}{24 \pi} \, \partial_+^3 \zeta
\;,
\label{452}\eea
while 
the anomalous variation of the $--$ component of the 2d CFT stress tensor vanishes. Here, $\kappa_3^2$ denotes Newton's constant in three
dimensions, while $c$ denotes the central charge associated with $\tau_{++}$,
\bea
\frac{c}{24 \pi}  = \frac{ \kappa_2^2}{\kappa_3^2}\,  \sqrt{2} \sqrt{G_4} \, p_L^3 \;.
\eea
Using $\kappa_3^2 = 2 \pi \, R_5 \kappa_2^2 $, see (4.10) in \cite{Cvetic:2016eiv}, we get, 
\bea
\frac{c}{24 \pi}  = \frac{\sqrt{2}}{2 \pi}  \frac{\sqrt{G_4}}{R_5} \, p_L^3 =  \frac{\sqrt{2}}{12 \pi}  \frac{\sqrt{G_4}}{R_5} \, c_L \;,
\eea
where the left moving central charge is \cite{Castro:2007sd}
\bea
c_L = 6 p_L^3 \;.
\eea
Thus
\bea
c =  2 \sqrt{2} \frac{\sqrt{G_4}}{  R_5} \, c_L \;.
\eea
 The central charge of the right moving sector \cite{Castro:2007sd} is completely encoded in the dimensionless 2d Newton's constant as
\be 
c_R = 6 \left(\frac{2}{\kappa_2^2}\right)^{3/2}.
\ee
 The above equation in tandem with \eqref{451} and \eqref{452} indicates that the putative boundary CFT$_1$ should be thought of as a chiral 
 CFT$_2$ ~\cite{Strominger:1998yg,Balasubramanian:2009bg}. 
 The chiral excitation temperature $T_L$ \cite{Castro:2007sd}  is determined in terms of $e^{-\psi}$ as
\bea
\pi \, R_5 \, T_L = \frac{2}{Y^0} = \frac{\kappa_2 \, e^{-\psi} }{\sqrt{2}} \;.
\eea

We note the similarity between the coefficient
of the anomalous variation given above in  \eqref{anomOsym}, and the coefficient of the anomalous variation given in 
\eqref{anomvarg4},
but we stress that they were obtained under different circumstances: in the former we kept $G_4$ fixed, while in the latter we varied $G_4$.

\section{Discussion} \label{sec_Discussion}
In this note, we consider 4d $\mathcal{N}=2$ Wilsonian effective actions with Weyl squared interactions.
These actions comprise gravity coupled to vector multiplets. By restricting to static spherically symmetric backgrounds, we perform a spherical reduction to obtain a 2d gravitational action. We restrict ourselves to constant scalar BPS backgrounds whose on-shell solutions correspond to AdS$_2$ space-times.  We first identify the counterterms that need to be added to the action to arrive at a well-defined variational principle. We then employ the holographic prescription to identify one-point functions of operators dual to the variations of the bulk metric, gauge fields and the composite field ln[$(Y^0)^2 \Upsilon$]. In the first two variations, we keep the four-dimensional Newton constant fixed, while in the latter we vary it. The corresponding operators are the trace of the boundary stress tensor ($\hat{\pi}^t_t$), operators with constant one-point functions proportional to the corresponding charges and the operator with one-point function ${\hat \Pi} $, respectively.

The stress tensor operator has both one- and higher-point functions vanishing, indicating that the on-shell asymptotically AdS$_2$ BPS backgrounds correspond to vacuum states of the dual theory, while the one-point function ${\hat \Pi}$  is a function of the time coordinate, implying that it undergoes a non-trivial variation under a general boundary coordinate transformation.  The trace of the holographic stress tensor operator $T$ induces diffeomorphisms in the boundary time coordinate, and the impact of the time variation on a generic operator $A$ in the boundary is captured by the correlator 
$\langle [T, A]  \rangle$. This leads to our first result: \newline

{\it The dimensionless operator $\frac{\hat{\Pi}}{2\sqrt{v_1}}$ in the dual CFT$_1$ has an anomalous variation under boundary diffeomorphisms, and the magnitude of the anomaly is precisely the $W^2$ corrected Wald entropy of the 4d BPS black hole, whose charges are encoded in the one-point functions of the boundary operators dual to the gauge fields.}\newline

In certain cases, the 4d black hole can be lifted up to five dimensions, and this anti-reduction process lifts up the AdS$_2$ background to a BTZ black hole in asymptotically AdS$_3$ geometry. According to the AdS$_3$/CFT$_2$ dictionary, BPS BTZ black holes are represented as an ensemble of chiral states in the dual CFT$_2$~\cite{Strominger:1997eq,Strominger:1998yg,Balasubramanian:2009bg}. At the 
two-derivative level, there are indications~\cite{Strominger:1998yg,Balasubramanian:2009bg}  that
the boundary CFT$_1$ dual to AdS$_2$ is the chiral half of the CFT$_2$. We assume this embedding to hold at the 4-derivative level and use it to determine how CFT$_2$ data relevant to computing BTZ entropy are encoded in the lower dimensional CFT.  These data include the central charge of the chiral sector $c_L$ that accounts for the entropy as well as the `chiral temperature' $T_L$, which is a measure of the excitation above the ground state that corresponds to the black hole microstates in the chiral sector. We begin by rewriting our bulk 2d background in `string frame' coordinates, which are the natural 2d coordinates one obtains via reduction from five dimensions. In this frame, we identify the CFT$_1$ one-point function
dual to the variation of the bulk 2d dilaton. The corresponding operator is a conformal dimension 2 operator with an 
anomalous transformation. At the two-derivative level, a precise identification of this operator with the chiral stress tensor of the CFT$_2$ has been written down in \cite{Cvetic:2016eiv}. Given this, we translate the result of the anomalous transformation of the dilaton operator expressed in 2d Newton's constant into units of 3d Newton's constant and arrive at our second result:
\newline

{\it The CFT$_1$ operator dual to the dilaton has an anomalous transformation whose anomaly is proportional to the central charge $c_L$ of the chiral half of the CFT$_2$ that encodes black hole entropy. }
\newline

The dimensionless chiral excitation temperature $T_L$ is expressed in terms of the boundary value of the 2d dilaton field as $\pi T_L R_5 = \frac{\kappa_2}{\sqrt{2}}e^{-\psi}$.
To sum up, in the 2d CFT picture, the black hole is represented as a chiral ensemble of states with a specific value of the excitation number $L_0 = |q_0|$ over the ground state with central charge $c_L$. However, there is no single operator  that encodes its entropy.  By contrast, in the 1d picture, the 
one-point function $\hat{\Pi}$ encodes the Wald entropy as the charge of its anomalous variation, though the chiral temperature $\sqrt{\frac{L_0}{c_L}}$ representing the excitation is read off from the dual 2d bulk theory as the on-shell value of the dilaton $e^{-\psi}$.

\vskip 5mm 

\subsection*{Acknowledgements}
We would like to thank Daniel Grumiller
for useful discussions. 
This work was partially
supported by FCT/Portugal through UIDB/04459/2020, through the LisMath PhD fellowship PD/BD/128415/2017 (P. Aniceto) and through the FCT Project CERN/FIS-PAR/0023/2019. 


\appendix

\section{Special geometry and Wilsonian Lagrangian \label{appendix_Special_geometry}}

The four-dimensional Wilsonian Lagrangian describing the coupling
of Abelian vector multiplets to $\mathcal{N}=2$ supergravity in the presence of Weyl squared interactions is encoded in a holomorphic
function $F(X, \hat A)$ that depends on complex scalar fields $X^I$ and $\hat A$ \cite{deWit:1996gjy}.
This function is 
homogeneous of second degree under scalings by $\lambda \in \mathbb{C} \backslash \{0\}$, i.e.
$F(\lambda X, \lambda^2 \hat A) = \lambda^2 F(X, \hat A)$. This implies the homogeneity relations
\bea
2 F &=& F_I  X^I + 2 {\hat A} F_{A} \nonumber\\
F_I &=& F_{IJ} X^J + 2 {\hat A} F_{{A} I} \;, \nonumber\\
F_{IJK} X^K &=& - 2 {\hat A} F_{A IJ} \;, \nonumber\\
2 {\hat A}  \, F_{A A } &=& - F_{A I } X^I \;, \nonumber\\
2 {\hat A} \,  F_{A A I } &=&  - F_{A I J} X^J - F_{A I} \;,
\label{sgrel}
\eea
where $F_I = \partial F (X, \hat A)/\partial X^I$ and $F_{A} =  \partial F (X, \hat A)/\partial \hat A$. Similarly, 
$F_{{A} I} =  \partial^2 F (X, \hat A)/\partial {\hat A}  \partial X^I$, 
 $F_{IJ} = \partial^2 F (X, \hat A)/\partial X^I \partial X^J$, and so on. 
 We denote the derivatives of ${\bar F} ( {\bar X}, \bar{\hat A}) $ with respect to ${\bar X}^I$ and $\bar{\hat A}$ by ${\bar F}_I$ and ${\bar F}_A$, respectively.
 Under the above scaling, $F_A$ has weight $0$,
 and so do $F_{IJ}$ and the combinations $N_{IJ} = -i (F_{IJ} - {\bar F}_{IJ} ), R_{IJ} = F_{IJ} + \bar{F}_{IJ} $. The following functions are symplectic
 scalars \cite{deWit:1996gjy},
 \bea
 F_A \;\;\;,\;\;\; F_{A I} N^{IJ} \left( F_J - {\bar F}_{JK} X^K \right) = - 2i {\hat A} \, \left( F_{AA} + i F_{A I } N^{IJ} F_{ A J} \right) \;.
 \label{symf}
 \eea

We refer to \cite{deWit:1996gjy,Cardoso:2006xz} for the full expression of the four-dimensional Wilsonian $\mathcal{N}=2$ supergravity Lagrangian
in the presence of Weyl squared ($W^2$) interactions, and for a brief description of the various supergravity fields entering in the Lagrangian.
Since we are interested in the class of solutions \eqref{2dBPS}, it is consistent to 
impose constraints on some of the supergravity fields.  One such constraint is to take the complex scalar fields $X^I$ and $\hat A$ to be constant.
Imposing these constraints, the four-dimensional $\mathcal{N}=2$ supergravity Lagrangian takes the form 
$ 8 \pi \, L_4=\mathcal{L}_1+\mathcal{L}_2$ with\footnote{
$\chi$ denotes the hyper-K\"ahler potential. $\hat C$ contains Weyl squared terms.}
 \cite{Cardoso:2006xz}
\begin{eqnarray}
  \label{eq:efflagwils}
    {\cal L}_1 &=&
    \ft14 i F_{IJ} F^{-I}_{ab} 
  (F^{-Jab} -\ft12 \bar X^J T^{- ab} )
   -\ft18 i F_I\,F^{+I}_{ab}
   \, T^{+ ab}
   + \tfrac12 i \hat F^{-ab}\, F_{{ A}I} \,F^{-I}_{ab} 
   + {\rm h.c.} \;,  \nonumber \\[1mm]
    {\cal L}_2 &=&   \mathrm{e}^{-\mathcal{K}} 
  (D- \ft16 R) +\ft12 \chi (D + \ft13 R)  
   -\ft1{32} \Big[ i (F-F_I X^I + \tfrac12 \bar F_{IJ} X^IX^J)
  (T_{ab}^+)^2
  + {\rm h.c.}\Big] 
  \nonumber\\
  &&
  +\ft12 \Big[ i F_{A} \hat C +\ft12 i F_{ A A}\, \hat
  F^-_{ab}\hat F^{-ab}   -\tfrac14 i \hat F^{-ab}\, F_{{
  A}I}\bar X^I  
  T_{ab}^-
  + {\rm h.c.}\Big]  \;,
\end{eqnarray}
where 
\begin{eqnarray}
\hat A &=& (T_{ab}^-)^2
 \;, \nonumber\\
 \mathrm{e}^{-\mathcal{K}} &=& i \left( {\bar X}^I F_I - {\bar F}_I X^I \right) \;.
 \end{eqnarray}
 In a spherically symmetric background, such as \eqref{backgr}, it is consistent to set
\begin{equation}
    T_{\underline{r}\underline{t}}^-
   =- i \, 
  T_{\underline{\theta}\underline{\varphi}}^-
 = w \;,
\end{equation}
where ${\underline{r}\underline{t}}, {\underline{\theta}\underline{\varphi}}$ denote Lorentz indices, and where $w$ denotes a complex scalar.
In addition, for the purpose of obtaining the class of solutions \eqref{2dBPS}, it is consistent to set the following fields to zero \cite{Cardoso:2006xz},
\begin{eqnarray}
D + \ft13 R = 0 \;, \nonumber\\
\hat{F}_{ab}^- = 0 \;.
\end{eqnarray}
The Lagrangian then simplifies to
\begin{eqnarray}
  {\cal L}_1 &=&
   \ft14 i F_{IJ} F^{-I}_{ab} 
  (F^{-Jab} -\ft12 \bar X^J T^{- ab} )
  -\ft18 i F_I\,F^{+I}_{ab}
  \, T^{+ ab}{}
  + {\rm h.c.}\;,   \\[1mm]
   {\cal L}_2 &=&   \mathrm{e}^{-\mathcal{K}} 
  (- \ft12 R) 
   -\ft1{32} \Big[ i (F-F_I X^I + \tfrac12 \bar F_{IJ} X^IX^J)
  (T_{ab}^+ )^2 + {\rm h.c.}\Big] 
   +\ft12 \Big[ i F_{A} \hat C 
  + {\rm h.c.}\Big]  \;.  \nonumber
\end{eqnarray}
Evaluating this Lagrangian in the background \eqref{backgr} and integrating 
over the 2-sphere, $ \tfrac12 \, {\cal F} = \int_{S^2} d \theta d \phi \, \sqrt{ - g_4} \, L_4$, gives the reduced Lagrangian
${\cal F} = {\cal F}_1 + {\cal F}_2$ given in \eqref{eq:F1-F2}.
In \eqref{eq:F1-F2} we expressed the reduced
Lagrangian in terms of the (rescaled) complex scalars $(v_2, Y^I, \Upsilon)$, but we can of course also express it
in terms of the set $(v_2, X^I, \hat A)$.

\section{AdS$_2$ in various coordinate systems \label{sec:ads2_sol}}

We summarize various coordinate systems for describing AdS$_2$ space-times.

The following line element describes 
AdS$_2$ in global coordinates~\cite{Sen:2011cn}
\bea
ds_2^2 = \frac{ a^2}{\sin^2 \sigma} \;
\left( - dT^2 + d\sigma^2 \right) \;\;\;,\;\;\; - \infty < T < \infty \;\;\;,\;\;\;   - \pi <  \sigma < 0 \;,
\label{globalads}
\eea
where $a^2 \in \mathbb{R}^+$.
Global 
AdS$_2$ has two boundaries, which in the above coordinates are located at  $\sigma = - \pi, 0$.

The line element 
\bea
ds_2^2 =a^2 \Big(  - r^2 dt^2 + dr^2/r^2 \Big) \;\;\;,\;\;\; 0 < r < \infty \;\;\;,\;\;\; -\infty < t < \infty \;.
\label{ads2poinc}
\eea
describes a patch of AdS$_2$, called Poincar\'e patch.
 The time-like Killing vector $\partial/\partial t$ becomes null at $r =0$.

The following line element also describes a space-time that is locally AdS$_2$,
\bea
ds_2^2 = a^2 \Big(- (\rho^2 -c^2) d\tau^2 + d\rho^2/(\rho^2 - c^2) \Big) \;\;\;,\;\;\; \rho > c > 0 \;.
\label{line2dbh}
\eea
This metric is referred to as 
describing the near-horizon geometry of a near-extremal black hole~\cite{Sen:2011cn}.
In these coordinates,
the outer Killing horizon is at $\rho=c > 0$.

All these line elements are constant curvature line elements
and can, locally, be brought into the form 
\bea
ds^2_2 = a^2 \Big( d {\tilde r}^2 - \left(  {\alpha} \, e^{\tilde r} + {\beta}  \, e^{-{\tilde r}} \right)^2 \, d {\tilde t}^2  \Big)
\label{fgline}
\eea
by a coordinate transformation. For instance, 
the line element \eqref{ads2poinc} can be brought into the form \eqref{fgline} with $\alpha = 1, \beta = 0$ by
the coordinate transformation $r = e^{\tilde r}$, while the line element \eqref{line2dbh} with $c=2$ can be 
brought into the form \eqref{fgline} with $\alpha = - \beta = 1$ by the coordinate transformation
\bea
\tilde r = \frac12 \ln \left(  \frac{ \rho + \sqrt{\rho^2 - 4} }{\rho - \sqrt{\rho^2 - 4} } \right) \;\;\;,\;\;\; \rho > 4 \;.
\eea

\section{Three ways of computing the Wald entropy of BPS black holes  \label{sec:Wald_entropy}}

We briefly describe three ways of computing the Wald entropy of BPS black holes.

First, consider a four-dimensional BPS black hole with a near-horizon geometry 
given by 
\bea
ds_4^2 &=& ds_2^2 + v_2 \, d\Omega_2^2 \;, \nonumber\\
ds_2^2 &=& v_1 \left( - r^2 dt^2 + dr^2/r^2  \right)\;,
\label{4dads2}
\eea
with $v_1 = v_2 > 0 $. The Killing horizon is located at  $r=0$, where the 
Killing vector $\xi = \partial/ \partial t$ becomes null.
The entropy of a static BPS black hole is computed using Wald's entropy formula in four dimensions \cite{Wald:1993nt,Iyer:1994ys,Jacobson:1994qe,Iyer:1995kg},
\bea
{\cal S}_{\rm Wald} \propto \int_{S^2} \frac{ \partial L_{4} }{\partial R_{abcd}} \epsilon_{ab} \epsilon_{cd} \;,
\eea
where $\partial L_{4}  / \partial R_{abcd}$ denotes the functional derivative of the Lagrangian $L_4$ with respect to the Riemann tensor $R_{abcd}$, with
the metric and the connection held fixed, 
where the Levi-Civita symbol $\epsilon_{ab}$ runs over Lorentz indices $a, b = {\underline{r}, \underline{t}}$, and where the integral is over 
a spatial cross-section of the Killing horizon, which in this case is $S^2$. Upon inclusion of the appropriate normalization factor, this results in
\cite{LopesCardoso:1998tkj}
\bea
{\cal S}_{\rm Wald}/\pi  =   i (\bar Y^IF_I -Y^I\bar F_I) 
+128 i (F_{\Upsilon} - \bar{F}_{\Upsilon} )   \;,
\label{4dbpsw}
\eea
with the horizon values $Y^I$ and $\Upsilon$ given by \eqref{BPSeq}.

Next, consider the near-horizon geometry of a near-extremal black hole in two dimensions, 
\bea
ds_2^2 = v_1 \Big[ - (\rho^2 -1) d\tau ^2 + d\rho^2/(\rho^2 -1) \Big]\;,
\label{2dbhv}
\eea
with $v_1>0$.
This describes a two-dimensional  black hole with an outer horizon at $\rho=1$, 
where the Killing vector  $\xi = \partial/ \partial \tau$ becomes null.
Its entropy 
 is computed using Wald's entropy formula in two dimensions,
\be
{\cal S}_{\rm Wald}= 2\pi 
\frac{ \partial L_{2} }{\partial R_{abcd}} \epsilon_{ab} \epsilon_{cd} 
 = \pi \left[- P'(R_2)\right]  \epsilon^{ab} \epsilon_{ab}
 = 2 \pi P'(R_2) \,,
 \label{wald2dbh}
\ee
where $L_2$ denotes the two-dimensional bulk Lagrangian given in \eqref{bulkr2lag2}, i.e. $\tfrac12 \, H = \sqrt{-h_2} \,  L_2$, and 
where $\epsilon^{ab} \epsilon_{ab} = -2$. When evaluated on a BPS solution, i.e. $v_2 R_2 = 2, \; \sqrt{-\Upsilon} = 8$, this yields 
\bea
{\cal S}_{\rm Wald}/\pi =  \, 2 P' (R_2) \vert_{BPS} =   i (\bar Y^IF_I -Y^I\bar F_I) 
+128 i (F_{\Upsilon} - \bar{F}_{\Upsilon} )   \:,
\label{bpswalde}
\eea
with the horizon values $Y^I$ given by \eqref{BPSeq}.
This reproduces the four-dimensional BPS entropy \eqref{4dbpsw}. Note, however, that the Killing vectors with respect
to which one calculates the Wald entropy are different, namely 
$\partial/ \partial t$ in four dimensions, and $ \partial/ \partial \tau$ in two dimensions. 

Thirdly, the two-dimensional entropy \eqref{wald2dbh} can also be obtained from a boundary computation, as follows.
We bring the line element \eqref{2dbhv}
of the two-dimensional black hole into the form \eqref{metfeff}, with the outer horizon located at $r_*$.
While in the main part of the paper we considered the boundary Lagrangian in \eqref{bbppaar2} at $r = \infty$, now we consider the boundary Lagrangian at 
the outer horizon $ r_*$,
since the Wald entropy of a two-dimensional black hole 
can be also be computed by resorting to this boundary Lagrangian, as follows \cite{Iyer:1995kg}.
We define the one-form density $\mathbf{B}$ to be 
\be
 \int \mathbf{B} = \tfrac12 \, S_1 -  \tfrac12 \, \int dt \left(\pi_I  \; \,, \; \tilde{\pi}^I \right)
\, \begin{pmatrix}
	A^I_t  - A^{ren \, I}_t  \\
	{\tilde A}_{I \, t}   - {\tilde A}^{ren}_{I \, t} 
\end{pmatrix} \,.
\ee
Then, according to eq. (56) in \cite{Iyer:1995kg}, the Wald entropy is
\be
{\cal S}_{\rm Wald}= 2 \pi \lim_{\epsilon \rightarrow 0} \int_{\mathcal{H}_\epsilon} \xi \cdot \mathbf{B} \,,
\ee
with $\xi^\mu \partial_\mu $ the horizon Killing vector field with unit surface gravity,  and where for a given 
hypersurface $\Sigma$ transverse to the vector field $\xi$, we define $\mathcal{H}_\epsilon$ as a smooth one-parameter family of surfaces in $\Sigma $ which approach the bifurcation surface (here a point) of the horizon $\mathcal{H}$ when $\epsilon \rightarrow  0$. 
In the coordinates \eqref{metfeff}, the Killing vector is $\xi = a \, \partial_t$, with $a$ a constant that is chosen so as to obtain unit surface gravity  $\kappa$ at
the horizon. Computing the 
surface gravity,
\bea
\kappa^2 = -\tfrac12 \left( \nabla^{\mu} \xi^{\nu} \right)  \left( \nabla_{\mu} \xi_{\nu} \right) \vert_{r_*} \;,
\eea
and using $a=  \frac{1}{\partial_{r} \sqrt{-\gamma}} \vert_{r_*} $ as well as 
$K = \partial_{r} \log \left(\sqrt{-\gamma}\right)$, we obtain for the Wald entropy of the BPS black hole, 
\bea
{\cal S}_{\rm Wald} &=&  \pi \,\lim_{r \rightarrow r_*} \frac{1}{\partial_{ r} \sqrt{-\gamma}} \left[ S_1 - \left(\pi_I  \; \,, \; \tilde{\pi}^I \right)
\, \begin{pmatrix}
	A^I_t  - A^{ren \, I}_t  \\
	{\tilde A}_{I \, t}   - {\tilde A}^{ren}_{I \, t} 
\end{pmatrix}    \right] \nn \\
 &=& 2\pi \, \lim_{r \rightarrow r_*}  \left[  \frac{\sqrt{-\gamma}}{\partial_r \sqrt{-\gamma}}  P'(R_2) \, K  \right]  \nn \\
 &=& 2 \pi P'(R_2) \vert_{BPS}\,,
\eea
in agreement with \eqref{bpswalde}.

\section{The Faddeev-Jackiw formalism applied to two-dimensional Einstein-Maxwell-dilaton theory \label{sec:HS}}

In this appendix we review the results presented in~\cite{Hartman2008dq} and improved in~\cite{Castro_Song_2014ima}, 
which describe an attempt at recasting 2d gravity as a CFT, whose boundary preserving gauge transformations give rise to a twisted stress-energy tensor with nonzero central charge. Here
we obtain the Dirac brackets of the theory considered in these papers using the Faddeev-Jackiw symplectic formalism~\cite{Faddeev:1988qp,Jackiw:1993in} as a way of circumventing the application of the Dirac-Bergmann algorithm. 
We briefly review this formalism and we apply it to the theory considered in~\cite{Hartman2008dq}, also allowing for the presence of a 
non-local Polyakov term in the action.

\subsection{Faddeev-Jackiw formalism}  \label{Appendix_Faddeev_Jackiw}
%

Here we present some details of the formalism used to compute the Dirac brackets for the theories we will be considering in the remainder of this appendix. By gauge fixing the Lagrangian in a conformal gauge we end up with a theory that is first order with respect to the fixed time derivative we choose. Consequently, we strictly focus on the specific case in which the Lagrangian is first order in time derivatives. This usually leads to the necessity of introducing constraints and applying the Dirac-Bergmann algorithm~\cite{dirac_1950,Bergmann_Anderson,Bergmann_Goldberg} to obtain the dynamics of the theory. Equivalently, one may follow the Faddeev-Jackiw symplectic formalism~\cite{Faddeev:1988qp,Jackiw:1993in} to arrive at the same commutation relations. To do so, we start by considering the tautological 1-form $\vartheta = \vartheta_i \delta \xi^i$.
  This 1-form can be read directly from the Lagrangian density of the theory, since for a field theory with a first order Lagrangian density we have
\be
\mathcal{L} = \vartheta_i\left(\xi \right) \dot{\xi}^i - V \left(\xi \right) \,,
\ee
up to total derivatives. Taking the standard Legendre transform we may identify $V$ with the Hamiltonian density $\mathcal{H}$ of the theory. The Euler-Lagrange equations yield
\be \label{eq_appendix_b_EL}
\left( \frac{\delta  \vartheta_j(\mathbf{x})}{\delta \xi^i (\mathbf{y})} - \frac{\delta  \vartheta_i (\mathbf{y})}{\delta  \xi^j  (\mathbf{x})}  \right) \dot{\xi}^j  (\mathbf{x})= \omega_{ij}(\mathbf{y},\mathbf{x})  \dot{\xi}^j  (\mathbf{x}) = \frac{\delta V (\mathbf{x})}{\delta \xi^i (\mathbf{y})} \,,
\ee
where we have defined the symplectic 2-form $\omega$ as
\be \label{eq_def_symplectic_2_form}
\omega = \delta \vartheta = \delta \int d\mathbf{x}  \, \vartheta_i(\mathbf{x}) \delta \xi^i(\mathbf{x}) = \frac{1}{2} \int d \mathbf{x} \, d \mathbf{y} \, \omega_{ij} (\mathbf{x}, \mathbf{y}) \delta \xi^i (\mathbf{x}) \wedge \delta \xi^j (\mathbf{y}) \,. 
\ee
When $\omega_{ij}$ is invertible\footnote{Here we will not be concerned with the more general case of non-invertible $\omega_{ij}$~\cite{Faddeev:1988qp,Jackiw:1993in,BarcelosNeto1991kw}.}
with its inverse denoted by $\omega^{ij}$ we can apply it to Eq.~\eqref{eq_appendix_b_EL} to obtain the evolution equation
\be
\dot{\xi}^i (\mathbf{z})  =\int d\mathbf{x}\, d \mathbf{y} \, \omega^{ij}(\mathbf{z},\mathbf{y})  \frac{\delta V (\mathbf{x})}{\delta \xi^j (\mathbf{y})} 
= \int d\mathbf{x}\, d \mathbf{y} \left[ \xi^i \left(\mathbf{z}\right)  , \xi^j  \left(\mathbf{y}\right) \right]  \frac{\delta V (\mathbf{x})}{\delta \xi^j (\mathbf{y})} \;,
\ee
with $\left[ \cdot  , \cdot  \right]  $ denoting the Dirac brackets for this theory.
By identifying $V= \mathcal{H}$, the above equation is seen to be the Hamiltonian equation of motion
\be \label{eq_appendix_b_eom_hamiltonian}
\dot{\xi}^i (\mathbf{x}) \equiv \left[\xi^i \left(\mathbf{x}\right)   , H   \right]  = \int d \mathbf{y} \,  \left[ \xi^i \left(\mathbf{x}\right)   , \mathcal{H} \left( \mathbf{y}\right)   \right]    \;.
\ee

\subsection{Application to the Hartman-Strominger theory} 
%
We now focus on the 2d action studied in~\cite{Hartman2008dq} and also discussed in~\cite{Castro_Song_2014ima}. This action is determined in terms of a metric $g_{\mu \nu}$, a gauge field $A_\mu$ and a scalar field $\eta$. The action (setting $G_2=c=1$) is given by
\be \label{eq_EMd_action_4d}
S= \frac{1}{2\pi}\int d^2t \, \sqrt{-g} \left[ \eta \left(R+ \frac{8}{l^2}\right)  - \frac{l^2}{4} F^{\mu \nu} F_{\mu \nu} 
\right] \,,
\ee
where $g$ is the determinant of the metric, $R$ its Ricci scalar and $F= dA$. Following~\cite{Hartman2008dq} we introduce an auxiliary field $f$ to substitute the quadratic gauge field term obtaining
\be \label{eq_EMd_action_with_aux}
S= \frac{1}{2\pi} \int d^2t \sqrt{-g} \left[\eta \left(R+\frac{8}{l^2}\right) - \frac{2}{l^2}f^2 + f \epsilon^{\mu \nu }F_{\mu \nu }\right]\,, 
\ee
with $\epsilon^{\mu \nu}$ the Levi-Civita tensor. Note that the $f$ equation of motions sets
\be
f= \frac{l^2}{4} \epsilon^{\mu \nu} F_{\mu \nu }\,.
\ee

To study the action of the 2d conformal group on this theory we follow the procedure detailed in~\cite{Hartman2008dq}. We rewrite it in the form of a standard 2d CFT. To do so, we take the conformal gauge for the metric
\be \label{eq_metric_CFT_reform}
ds^2 = - e^{2\rho} dt^+ dt^- \,.
\ee
This fixes the metric locally up to conformal diffeomorphisms generated by $\zeta^+(t^+)$ and $\zeta^-(t^-)$. Furthermore, we choose the Lorenz gauge for the gauge field
\be \label{eq_Lorenz_gauge}
\partial_+ A_- + \partial_- A_+  = 0 \,,
\ee
which fixes the field up to residual transformations generated by $\theta(t^+)+ \tilde{\theta}(t^-)$. In the gauge \eqref{eq_Lorenz_gauge} $A$ is determined in terms of a scalar field, 
\be \label{eq_gauge_fixed_A}
A_\pm = \pm \partial_{\pm} a  \implies F_{+-} = -2 \partial_- \partial_+ a \,.
\ee
The action, up to boundary terms, becomes
\be  \label{eq_CFT_action}
I= \frac{1}{2\pi } \int dt^2 \left(2 \partial_\mu f \partial^\mu a -2 \partial_\mu\eta \partial^\mu \rho  + \frac{4}{l^2} e^{2\rho} \eta - \frac{1}{l^2} e^{2\rho}f^2\right) \,,
\ee
where we are now raising and lowering indices with respect to the metric $ds^2 = 2 dt^+ dt^-$. As mentioned in~\cite{Hartman2008dq,Castro_Song_2014ima}, the equations of motion obtained from this action must be supplemented by the gauge and gravitational constraints coming from~\eqref{eq_EMd_action_with_aux}. Defining 
\be
T_{\pm \pm} = - \frac{2\pi }{\sqrt{-g}} \frac{\delta I}{\delta g^{\pm \pm}} \,,\qquad j_{ \pm }= -2\pi  \frac{\delta I}{\delta A_\mp} \,, 
\label{tpm}
\ee
the constraints enforce
\be
j_{\pm} = \pm 2 \partial_\pm  f = 0 \,, \qquad T_{\pm \pm } = \partial^2_\pm \eta- 2 \partial_\pm \eta \partial_\pm \rho  + j_\pm A_\pm= 0 \,,\label{eq_T__constraint}
\ee
where we have added the term $j_\pm A_\pm$ to $T_{\pm \pm}$, so that the latter is holomorphically conserved without use of the constraints~\cite{Hartman2008dq}.

We proceed with the computation of the Dirac brackets of this theory at fixed $t^+$ to show how these constraints generate the residual gauge and diffeomorphism  transformations. For the Lagrangian defined in~\eqref{eq_CFT_action}  the tautological one-form $\vartheta$ corresponds, up to boundary terms, to 
\be \label{eq_tautological_1_form}
\vartheta(t^-) = \frac{2}{\pi} \left[ f'(t^-) \delta a(t^-) - \eta'(t^-) \delta \rho(t^-)  \right] \,,
\ee
where $'$ corresponds to derivatives with respect to $t^-$. Applying~\eqref{eq_def_symplectic_2_form} to~\eqref{eq_tautological_1_form} we obtain
\be \label{eq_omega_matrix}
\left[\omega_{ij}(t^-,s^-)\right] = \begin{bmatrix} \begin{array}{c|ccccccc}
		& \rho&\eta  & a& f   \\
		\hline
		\rho & 0 & 1& 0  & 0  \\
		\eta &  1& 0 & 0 & 0   \\
		a &  0 & 0 & 0 & -1  \\
		f &  0 & 0 & -1 & 0   
	\end{array}
\end{bmatrix}  \frac{2}{\pi}  \delta'( t^-- s^-) \,.
\ee
We may then invert this matrix to obtain the Dirac brackets
\be \label{eq_Dirac_brackets}
\left[\eta(t^-), \rho (s^-)\right] = \frac{\pi}{4} \mathrm{sgn}(t^- - s^-)  \,, \qquad \left[a(t^-), f(s^-)\right] = - \frac{\pi}{4} \mathrm{sgn}\left(t^-- s^-\right)\,,
\ee
with $\mathrm{sgn}$ denoting the sign function.

The equations of motion are obtained through Eq.~\eqref{eq_appendix_b_eom_hamiltonian} where $H$, the Hamiltonian, is given by
\be 
H= \frac{1}{2\pi l^2} \int ds^- \, e^{2\rho} \left(f^2 - 4 \eta \right) \,.
\ee

The Dirac brackets in~\eqref{eq_Dirac_brackets} can be used to determine the commutation relations of $j_\pm$ and $T_{\pm \pm}$. At fixed $t^+$ these are
\begin{align}
\label{eq_commutator_stress_energy_tensors}
\frac{1}{\pi}\int ds^- \left[T_{--}(t^-),\zeta(s^-) T_{--}(s^-)\right] &=  \zeta(t^-) \partial_{t^-}T_{--} (t^-) + 2 \partial_{t^-} \zeta(t^-) T_{--}(t^-) \,, \\
\frac{1}{\pi}\int ds^- \left[j_{-}(t^-), \zeta(s^-) T_{--}(s^-)\right] &=  \zeta(t^-) \partial_{t^-}j_-(t^-) + \partial_{t^-} \zeta(t^-)  j_-(t^-) \,, \\
\frac{1}{\pi}\int ds^- \left[T_{--}(t^-), \theta(s^-) j_{-}(s^-)\right] &= j_-(t^-)  \partial_{t^-} \theta(t^-)  \,, \\
\frac{1}{\pi}\int ds^- \left[A_-(t^-), \theta(s^-) j_{-} (s^-)\right] &= \partial_{t^-} \theta(t^-)  \,. \label{eq_commutator_gauge_transformation}
\end{align}
Eqs.~\eqref{eq_commutator_stress_energy_tensors}-\eqref{eq_commutator_gauge_transformation} show that $T_{\mp \mp}$ and $j_\mp$ are respectively the generators of the residual diffeomorphism and gauge transformations (at fixed $t^\pm$).

As shown in~\cite{Hartman2008dq}, the asymptotic symmetry transformations correspond to conformal diffeomorphisms that are
supplemented by a certain gauge transformation, and they are generated by the twisted stress-energy tensor
\be
\tilde{T}_{\pm \pm} = T_{\pm \pm} \pm \frac{E l^2}{4}  \partial_\pm j_\pm \,,
\ee
with $E$ given by the vacuum solution $F_{+-} = 2E \epsilon_{+-}$. For this twisted stress-energy tensor to transform anomalously, taking into account~\eqref{eq_commutator_stress_energy_tensors}, it is assumed in~\cite{Hartman2008dq} that the conserved currents $j_\pm$ are not gauge invariant and that under a gauge transformation they transform as
\be
\frac{1}{\pi} \int ds^- \, \left[ j_-\left(t^-\right) , \theta\left(s^-\right)  j_-\left(s^-\right) \right] = k \partial_{t^-} \theta\left(t^-\right) \,.
\ee 
However, it is clear from~\eqref{eq_Dirac_brackets} that the theory we are considering does not provide a way of obtaining this result, and in particular the level $k$ is still left undetermined. In~\cite{Castro_Song_2014ima} an argument is given for why $k \neq 0$ by making contact with the holographic duals of the metric and gauge field obtained through a specific holographic renormalization of the 2d theory. However, this holographic renormalization procedure is predicated on the usage of a boundary term introduced in~\cite{Castro:2008ms}  that is not gauge invariant, which as discussed in~\cite{Cvetic:2016eiv}, leads to pathologies in the dual theory, and a direct comparison with the bulk quantities is therefore not straightforward.

\subsection{Non-local Polyakov term} 

%
In this subsection we show that the addition of the non-local Polyakov term to the action \eqref{eq_EMd_action_with_aux}
can still be treated using the Faddeev-Jackiw symplectic formalism discussed above and that it
leads, as it should, to the appearance of  a central charge in the bulk theory. 

We add the following term to the action \eqref{eq_EMd_action_with_aux},
\be \label{eq_Polyakov_non_local}
S_{\mathrm{P}} = -\frac{c}{96\pi} \int d^2t \sqrt{-g} \left[ R \, \Box^{-1} R  \right]\,.
\ee
In conformal gauge (c.f.~\eqref{eq_metric_CFT_reform}) we obtain $\Box^{-1} R=-2\rho$, and therefore~\eqref{eq_Polyakov_non_local} after integrating by parts simplifies to~\cite{Callan:1992rs}
\be \label{eq_Polyakov_action_conformal_gauge}
S_\mathrm{P}  = -\frac{c}{12 \pi}  \int d^2t \, \partial_- \rho \partial_+ \rho  \,.
\ee
$T_{\pm \pm } $ in \eqref{tpm} receives the following
contributions from this new term, 
\be
T^\mathrm{P}_{\pm \pm } =-\frac{2\pi }{\sqrt{-g}} \frac{\delta S_\mathrm{P}}{\delta g^{\pm \pm}} = \frac{c}{12} \left[ \partial_-^2 \rho - \left(\partial_- \rho\right)^2  \right]   \,,
\ee
where we have used
\bea
&&-\frac{2\pi }{\sqrt{-g}}   \frac{\delta S_\mathrm{P}}{\delta g^{\mu \nu}}  = \\
&& \frac{c}{48}  \left[ 2g_{\mu \nu} R -2 \nabla_\mu \nabla_{\nu} \Box^{-1}R  + \nabla_\mu \left(\Box^{-1}R\right) \nabla_\nu \left(\Box^{-1}R\right) - \frac{g_{\mu \nu}}{2} \nabla_\lambda \left(\Box^{-1}R\right) \nabla^\lambda \left(\Box^{-1}R\right)\right] \,. \nonumber
\eea
Taking the trace of this expression we obtain $T^\mu_{\; \mu } = c R/24$.

The symplectic 2-form~\eqref{eq_def_symplectic_2_form} at constant $t^+$ also receives a modification, which can be read directly from \eqref{eq_Polyakov_action_conformal_gauge} 
\be \label{eq_omega_matrix_non_local_direct_treatment}
\omega_{\rho \rho}(t^-,s^-)=  \frac{c}{6\pi}  \delta'( t^- - s^-) \,.
\ee
Note that this term modifies~\eqref{eq_omega_matrix}. By inverting the modified $\omega_{ij}$ we find that the Dirac brackets of~\eqref{eq_Dirac_brackets} still hold, but in addition there is also a new non-trivial commutator given by
\be
\left[\eta(t^-), \eta(s^-) \right] = -\pi \frac{c}{48}  \sgn \left(t^- - s^-\right) \,.
\ee
This in turn implies
\bea \label{eq_Dirac_bracket_rho}
\frac{1}{\pi} \int ds^- \left[T_{--}(t^-), \zeta(s^-)T_{--}(s^-)\right] = \zeta (t^-) \, \partial_- T_{--} + 2 \partial_- \zeta (t^-) \,  T_{--} + \frac{c}{24} \partial^3_- \zeta (t^-)  \;.
\nonumber\\
\eea
%


\providecommand{\href}[2]{#2}\begingroup\raggedright\endgroup

\end{document}